\documentclass[english,11pt,a4paper]{article}
\pdfoutput=1 

\usepackage{jheppub,amsmath} 

\usepackage[T1]{fontenc}
\usepackage{color}
\usepackage{babel}
\usepackage{amstext}
\usepackage{amssymb}
\usepackage{graphicx}
\usepackage{esint}

\usepackage{times}
\usepackage{xspace}
\usepackage{caption}
\usepackage{subcaption}
\usepackage{slashed}
\usepackage{verbatim}
\usepackage[utf8]{inputenc}
\usepackage{breqn}

\def\be{\begin{eqnarray}}
\def\ee{\end{eqnarray}}

\newcommand{\mt}[1]{\textrm{\tiny #1}}

\def\Raa{R_{\mt{AA}}}

\newcommand{\pythia}{{\sc Pythia}\xspace}

\def\RAA{\Raa}

\newcommand{\diff}{\mathop{}\!\mathrm{d}}
\usepackage{siunitx}
\title{Evolution of the Mean Jet Shape and Dijet Asymmetry Distribution of an Ensemble of Holographic Jets in Strongly Coupled Plasma}

\author[a]{Jasmine Brewer,}
\author[a]{Krishna Rajagopal,}
\author[a,b]{Andrey Sadofyev,}
\author[a,c]{Wilke van der Schee}

\affiliation[a]{Center for Theoretical Physics, Massachusetts Institute of Technology, Cambridge, MA 02139, USA}

\affiliation[b]{Theoretical Division, MS B283, Los Alamos National Laboratory, Los Alamos, NM 87545}

\affiliation[c]{Institute for Theoretical Physics and Center for Extreme Matter and
Emergent Phenomena, Utrecht University, Leuvenlaan 4, 3584 CE Utrecht,
The Netherlands}

\emailAdd{jtbrewer@mit.edu}
\emailAdd{krishna@mit.edu}
\emailAdd{sadofyev@mit.edu}
\emailAdd{wilke@mit.edu}

\preprint{\begin{minipage}[t]{0.2\textwidth}\raggedleft\footnotesize{MIT-CTP-4943\\[-0.1em]
LA-UR-17-29843}\end{minipage}}

\abstract{
Some of the most important experimentally accessible probes of the quark-gluon plasma (QGP)
produced in heavy ion collisions come from the analysis of how
the shape and energy of sprays of energetic particles produced
within a cone with a specified opening angle (jets) 
in a hard scattering are modified by their passage through the
strongly coupled, liquid, QGP.
We model an ensemble of back-to-back dijets
for the purpose of gaining a qualitative understanding of how the shapes of the individual jets and 
the asymmetry in the energy of the pairs of jets in the ensemble are modified
by their passage through
an expanding cooling droplet of strongly
coupled plasma, in the model in a holographic gauge theory that is dual to a 4+1-dimensional 
black-hole spacetime that is asymptotically anti-de Sitter (AdS). 
We build our model by constructing an ensemble of strings in the
dual gravitational description of the gauge theory. We model QCD jets in vacuum
 using strings whose endpoints are moving ``downward'' into the gravitational bulk spacetime
with some fixed small angle, an angle that represents the opening angle (ratio of jet mass to jet energy) that the QCD jet would have in vacuum.  
Such strings must be moving through the gravitational bulk at (close to) the speed of light; they must be (close to) null. 
This condition does not specify the energy distribution along the string, meaning that it does not specify the shape of the jet being modeled.
We study the dynamics of strings that are initially not null and show that strings with a wide range of initial conditions rapidly accelerate and become null and, as they do, develop a similar distribution of their energy density.
We use this distribution of the energy density along the string, 
choose an ensemble of strings whose opening angles and energies are 
distributed as in perturbative QCD, and show that 
we can then fix one of the two model parameters such that
the mean jet shape for the jets
in the ensemble  that we have built matches that measured in proton-proton collisions
reasonably well.
This is a novel way for hybridizing relevant inputs from perturbative QCD and 
a strongly coupled holographic gauge theory in the service of modeling 
jets in QGP.
We send our ensemble
of strings through an expanding cooling droplet of strongly coupled plasma,
choosing the second model parameter so as to get a reasonable value for $\RAA^{\rm jet}$, the 
suppression in 
the number of jets, 
and study
how the mean jet shape and the dijet asymmetry are modified, comparing
both to measurements from heavy ion collisions at the LHC.
}

\begin{document}
\maketitle
\flushbottom

\section{Introduction}
\label{sec:intro}


Ultrarelativistic 
heavy ion collisions at the Relativistic Heavy Ion Collider (RHIC) and the Large Hadron Collider (LHC)  recreate droplets of the hot matter that filled the microseconds old universe, called quark-gluon plasma (QGP).   Experiments at these facilities provide unique experimental access to the
properties of QGP as well as to the dynamics via which droplets of QGP form, expand and cool.
These experiments 
have demonstrated that in the experimentally accessible range of temperatures, up to several times hotter than the crossover temperature at which cooling QGP becomes ordinary hadronic matter, 
droplets of QGP 
exhibit strong collective 
phenomena~\cite{Adcox:2004mh,Arsene:2004fa,Back:2004je,Adams:2005dq, Aamodt:2010pa,ATLAS:2011ah,Chatrchyan:2012ta},
with the dynamics of the rapid expansion and cooling of the initially lumpy droplets produced in 
the collisions
successfully described by the equations of  relativistic viscous hydrodynamics~\cite{Teaney:2000cw,Huovinen:2001cy,Teaney:2001av,Hirano:2005xf,Romatschke:2007mq,Luzum:2008cw,Schenke:2010rr,Hirano:2010je,Gale:2012rq,Shen:2014vra,Shen:2014nfa,Ryu:2015vwa,Bernhard:2016tnd,Bass:2017zyn}.
The ratio of the shear viscosity, $\eta$, to the entropy density, $s$, serves as a benchmark, because in a weakly coupled plasma, $\eta/s\propto  1/g^4$ (with $g$ the gauge coupling), meaning that this ratio is large, 
whereas
$\eta/s = 1/4\pi$ in the high temperature phase (conventionally called the plasma phase even though in reality it is a liquid) of any gauge theory that has a dual gravitational description in the limit of strong coupling and large number of colors~\cite{Policastro:2001yc,Buchel:2003tz,Kovtun:2004de}. Comparisons between hydrodynamic calculations of, and experimental measurements of, anisotropic flow in heavy ion collisions indicate that the QGP in QCD has an $\eta/s$ that is comparable to, and in particular not much larger than $1/4\pi$, meaning 
that QGP itself is a strongly coupled liquid.


The discovery that 
QGP is a strongly coupled liquid at length scales of order its inverse temperature and longer
even though (because QCD is asymptotically free) it consists of weakly coupled quarks and gluons when probed with high resolution
challenges us to find experimental means to probe QGP at multiple length scales.
The only probes that we have available are those
produced in the same heavy ion collisions in which the droplets of QGP themselves are produced.
Here we shall focus entirely on the use of high transverse momentum jets, produced at the moment of the collision in initial hard scatterings, as probes.
Jets are produced with some energy and virtuality, the latter 
often also referred to as the jet mass.  Assuming that the jet propagates in vacuum, both are (almost) conserved during the development and branching of the partonic jet shower that occurs after the jet is produced in an initial hard scattering. (Only almost because the jet may exchange soft momenta with the underlying event or with other jets.)  The partonic shower develops within a cone whose opening angle 
is proportional to the ratio of the jet mass to the jet energy.
As a partonic jet shower propagates through the strongly coupled
plasma created in a heavy ion collision, however, the partons in the shower each
lose energy and momentum as a consequence of their strong interactions 
with the plasma, creating a wake in the plasma.  
These interactions lead to a reduction in the  jet energy (or quenching) 
and to modifications  
of the opening angle and shape
of jets produced in heavy ion
collisions relative to those of their counterparts produced in proton-proton collisions, that propagate in vacuum.
By pursuing a large suite of jet measurements, the different LHC collaborations have observed 
strong modification of different jet observables  in heavy ion collisions~\cite{Aad:2010bu,Chatrchyan:2011sx,Chatrchyan:2012nia,Chatrchyan:2012gt,Chatrchyan:2012gw,Aad:2012vca,Raajet:HIN,Aad:2013sla,Chatrchyan:2013kwa,Abelev:2013kqa,Chatrchyan:2013exa,Chatrchyan:2014ava,Aad:2014wha,Aad:2014bxa,Adam:2015ewa,Adam:2015doa,Aad:2015bsa,HIN-16-002,Khachatryan:2015lha,Khachatryan:2016erx,Khachatryan:2016tfj,Khachatryan:2016jfl,Aaboud:2017bzv,Acharya:2017goa,Sirunyan:2017jic,Aaboud:2017eww},  
making jets promising QGP probes. 
The first experimental constraints on jet quenching came from hadronic measurements at RHIC~\cite{Adcox:2001jp,Adler:2002xw,Adler:2002tq}. Analyses of
jets themselves and their modification 
are also being performed at RHIC~\cite{Ploskon:2009zd,Perepelitsa:2013faa,Adamczyk:2013jei,Jacobs:2015srw,Adamczyk:2016fqm,Adamczyk:2017yhe} and are one of the principal scientific goals of the
planned sPHENIX detector~\cite{Adare:2015kwa}.


A complete theoretical description of the processes by which jets are modified via passage through QGP remains challenging for the same reason that it is interesting, namely 
because it is a multi-scale problem.
The production of jets and the processes via which an initial hard parton fragments into a shower
are weakly coupled hard processes.
However, the dynamics of the droplet of QGP including the wake produced in it by the 
passing jets and, more generally, the interaction of the jets with the QGP are sensitive
to strongly coupled physics at scales of order the temperature of the QGP.
One class of theoretical
approaches is based upon assuming that suitably resummed weakly coupled analyses can be
applied almost throughout.
(See Refs.~\cite{Jacobs:2004qv,CasalderreySolana:2007pr,Majumder:2010qh,Mehtar-Tani:2013pia,Ghiglieri:2015zma,Blaizot:2015lma,Qin:2015srf} for reviews. 
Based on these approaches, Monte Carlo tools for analyzing jet observables are being 
developed~\cite{Zapp:2008af,Zapp:2008gi,Armesto:2009fj,Schenke:2009gb,Lokhtin:2011qq,Zapp:2012ak,Zapp:2013vla,Zapp:2013zya,Cao:2017zih} 
and many phenomenological studies of jets in medium have been confronted with LHC measurements of a variety of jet observables~\cite{Vitev:2009rd,CasalderreySolana:2010eh,Qin:2010mn,Young:2011qx,He:2011pd,CasalderreySolana:2011rq,Renk:2012cx,Neufeld:2012df,Renk:2012cb,Dai:2012am,Apolinario:2012cg,Zapp:2012ak,Wang:2013cia,Ma:2013pha,Huang:2013vaa,Senzel:2013dta,Zapp:2013vla,Zapp:2013zya,Ramos:2014mba,Renk:2014lza,Perez-Ramos:2014mna,Chien:2015vja,Huang:2015mva,Chien:2015hda,Milhano:2015mng,Zhang:2015trf,Chang:2016gjp,Mueller:2016gko,Chen:2016vem,Mehtar-Tani:2016aco,Tachibana:2017syd,KunnawalkamElayavalli:2017hxo,MIlhano:2017nzm,Mehtar-Tani:2017web}.)
However, since QGP  is a strongly coupled liquid we know that 
physics at scales of order its temperature
must be 
governed by strong coupling dynamics.
This realization has 
opened the door to many 
connections between the physics of the 
QCD plasma 
and gauge/gravity duality~\cite{Maldacena:1997re}, which
yields rigorous and quantitative 
access to nonperturbative, strongly coupled, physics 
in a large family of non-abelian gauge theory plasmas
that have a dual holographic description in terms of a black hole spacetime in a gravitational theory with one higher dimension. 
The AdS/CFT correspondence has become very successful in recent years
for describing strongly-coupled dynamics in a variety of arenas. 
In its simplest form, AdS/CFT
provides a duality between strongly coupled $\mathcal{{N}}=4$ supersymmetric Yang-Mills (SYM) theory
in 3+1 dimensions and classical Einstein gravity in 4+1-dimensional AdS space, or a 4+1 dimensional black hole that is asymptotically AdS in the case where the ${\cal N}=4$ SYM theory is at a nonzero temperature.
Although this AdS/CFT duality has not been shown 
to apply to QCD, the study of the plasmas in gauge theories that do have a holographic description has led to many qualitative 
insights into the properties and dynamics of QGP. (See 
Refs.~\cite{CasalderreySolana:2011us,DeWolfe:2013cua,Chesler:2015lsa} for reviews.) 
Within this context, there have been many interesting studies that address varied aspects of the interaction between high energy probes and strongly 
coupled plasma~\cite{Herzog:2006gh,Liu:2006ug,CasalderreySolana:2006rq,Gubser:2006bz,Liu:2006nn,Liu:2006he,Gubser:2006nz,Chernicoff:2006hi,CasalderreySolana:2007qw,Chesler:2007an,Gubser:2007ga,Chesler:2007sv,Hofman:2008ar,Gubser:2008as,Hatta:2008tx,Dominguez:2008vd,Chesler:2008wd,Chesler:2008uy,DEramo:2010wup,Arnold:2010ir,Arnold:2011qi,Arnold:2011vv,Chernicoff:2011xv,Chesler:2011nc,Arnold:2012uc,Arnold:2012qg,Chesler:2013urd,Ficnar:2013wba,Ficnar:2013qxa,Chesler:2014jva,Rougemont:2015wca,Chesler:2015nqz,Casalderrey-Solana:2015tas,Rajagopal:2016uip,Brewer:2017dwd}.
No holographic analysis can --- by itself --- treat the intrinsically weakly coupled processes of jet production and fragmentation, since in all examples that are currently accessible via gauge/gravity duality the gauge theory is strongly coupled in the ultraviolet, rather than asymptotically free.



There are now two quite different phenomenological 
approaches being developed with the goal of
addressing the multi-scale dynamics of QCD jets in strongly coupled plasma more fully, blending inputs from perturbative QCD calculations and holographic calculations where each may be relevant.
The authors of
Refs.~\cite{Casalderrey-Solana:2014bpa,Casalderrey-Solana:2015vaa,Casalderrey-Solana:2016jvj,Hulcher:2017cpt} 
have developed a
hybrid strong/weak coupling model in which perturbative QCD parton showers taken from \pythia
are modified, parton-by-parton, 
upon assuming that
the interaction between each parton formed in the shower 
and the QGP
follows the rate of energy loss of an energetic quark in strongly coupled
plasma obtained via the holographic calculations 
in Refs.~\cite{Chesler:2014jva,Chesler:2015nqz}.  They have confronted their hybrid model
with various suites of experimental data and in so doing have obtained qualitative insights into the implications of measurements of jet suppression, jet shapes, jet fragmentation functions, and the suppression, energy asymmetry and angular distributions of dijets, gamma-jets and Z-jets for parton energy loss, transverse momentum broadening, the degree to which the wakes left  in the plasma
by passing jets have time to equilibrate, and the resolving power of QGP.

The second approach, which we shall further develop here, was introduced by three of 
us in Ref.~\cite{Rajagopal:2016uip} and is more ambitious in its use of holography, 
as we model each jet in its entirety as an energetic massless quark plowing
through the plasma of ${\cal N}=4$ SYM theory.
In holography, the dynamics of quarks in the fundamental representation
is studied by adding spacefilling D7 branes to the bulk spacetime.
Open strings can end anywhere within a D7 branes, and are dual to a quark-antiquark
pair in the dual boundary CFT \cite{Karch:2002sh}. 
These open strings can be constructed in many different kinds of configurations and have 
been used to model varied dynamical phenomena. As we shall discuss at greater length below,
a pair of light
quark jets in plasma is described by an open fundamental string whose endpoints
shoot away from each other and the same time fall ``downwards'' into the black hole in the additional
dimension in the AdS spacetime, with the downward angle of their motion representing (i.e.~being proportional to) the opening angle of the jet in the gauge theory. 
One way of looking at the approach to modeling jets introduced in 
Ref.~\cite{Rajagopal:2016uip} is that we seek to use inputs from perturbative
QCD that are in a sense minimal, namely only those inputs that describe jet production.
The way we do this is to construct an ensemble of holographic jets with an initial
probability distribution for their energy and opening angle taken from perturbative QCD
so as to reproduce this distribution as in proton-proton collisions.
The qualitative insight obtained in Ref.~\cite{Rajagopal:2016uip} is that
even though every jet in the ensemble widens as it propagates through
the strongly coupled plasma, after passage through the plasma jets 
with a given energy in the ensemble can have a smaller mean opening angle
than jets with that energy would have had if they were in vacuum.
This happens because there are far fewer jets with higher energies than with lower energies
in the distribution (before quenching the 
distribution is $\sim E_{\rm jet}^{-6})$ and because those jets that are initially wider lose more energy, meaning
that the jets that remain with any specified energy are those narrow jets which suffered the least energy loss.
This result highlights the importance of analyzing an ensemble of
jets if one wishes to make comparisons, even qualitative comparisons, to
jet phenomenology: because different jets with the same $E_{\rm jet}$ that
traverse the same plasma but that start out with different initial opening angles
lose very different amounts of energy, it is insufficient and in fact quite misleading
to attempt to draw phenomenological conclusions by looking just at single average jet with
some given energy.
This conclusion applies for very similar reasons in perturbative~\cite{Milhano:2015mng},  holographic~\cite{Rajagopal:2016uip},
and hybrid~\cite{Casalderrey-Solana:2016jvj} calculations.

In the present study, which we reported on preliminarily in Ref.~\cite{Brewer:2017dwd},
we extend the model of Ref.~\cite{Rajagopal:2016uip} in two important ways.
First, we analyze the shape of the jets in the ensemble, rather than just their opening angle.
This forces us to consider the initial distribution of energy along the string more carefully.
Our goal is to choose this distribution so as to reproduce the shape of jets in vacuum,
and then to study how this shape is modified by passage through the plasma.
In Section~\ref{sec:strings} we shall find a rather remarkable way of using quite nontrivial string dynamics in the 
holographic gauge theory to construct an ensemble of strings (in Section~\ref{sec:model}) 
whose mean jet shape
does indeed reproduce the mean shape of QCD jets produced in proton-proton collisions.
Second, we choose an ensemble of back-to-back dijets with the distribution of the 
energy asymmetry between the two jets in an event
chosen to match that 
measured in proton-proton collisions and analyze how this dijet asymmetry distribution
is modified by passage through the plasma.

\begin{figure}[t]
\begin{centering}
\includegraphics[width=14cm]{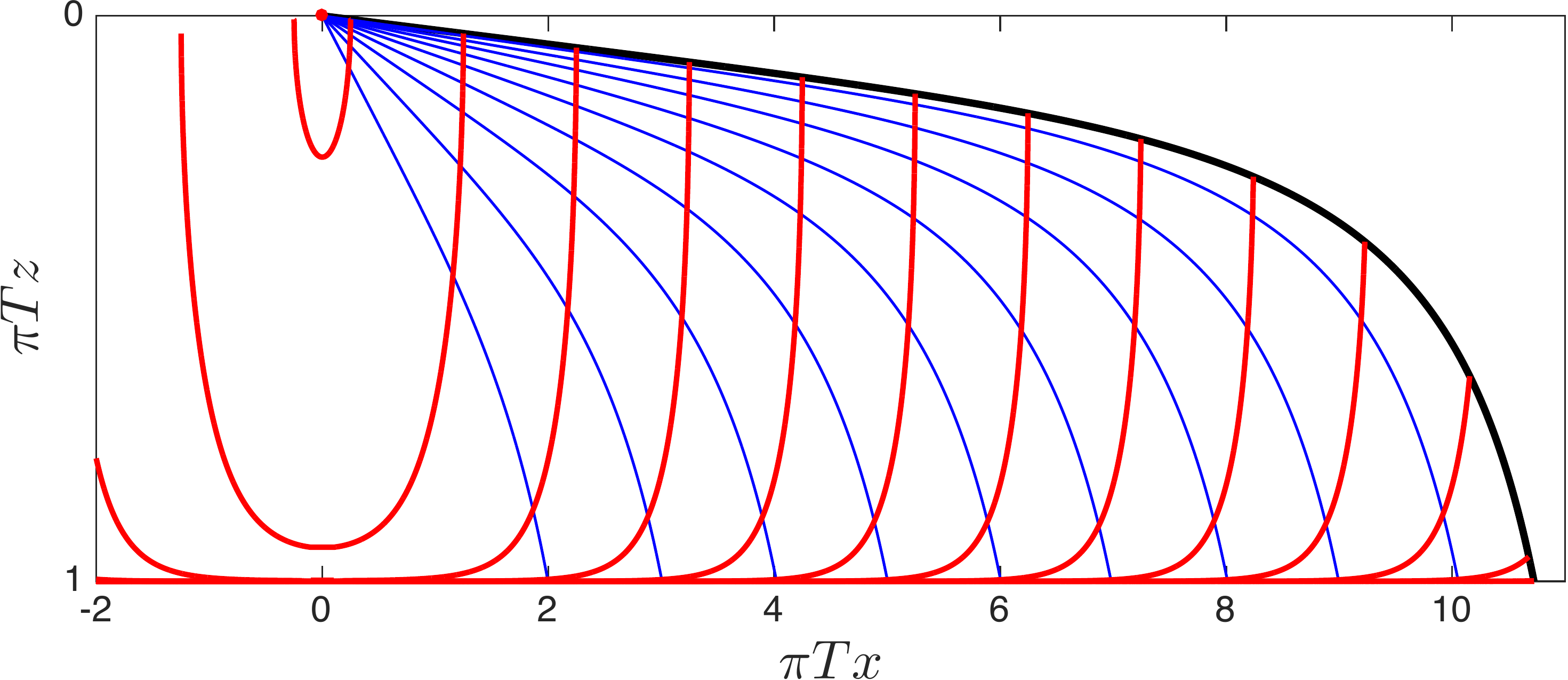}
\par\end{centering}
\protect\caption{
\label{fig:NullString} 
A null string (red) used to model a jet moving in the $x$-direction
shown at several coordinate times $t$~\cite{Chesler:2015nqz}.  
The string starts off at the point $x=0$ on the boundary and expands at the speed of light while falling downwards into the holographic $z$-direction, towards the horizon located at $z=z_h=1/(\pi T)$.
The blue curves represent the null geodesics that each (red) bit of energy that makes up the string follows;
the black curve is the endpoint trajectory.  
Different blue curves are parametrized by different values of $\sigma$, where $\sigma$ is the initial downward angle
in the $(x,z)$ plane.  The endpoint is the trajectory with $\sigma=\sigma_0$; 
the figure is drawn with $\sigma_0=0.025$.
If the string were in vacuum, there would be no horizon,
the blue null geodesics would all be straight lines, and the red string would maintain its initially semi-circular shape forever~\cite{Chesler:2008wd}.  
The opening angle of the jet which such a vacuum string represents
is proportional to $\sigma_0$.
Due to the presence of the
horizon, which is to say due to the presence of the strongly coupled plasma, a blue (or black) trajectory with
a given $\sigma$ curves downward: its angle in the $(x,z)$ plane, which starts out equal to $\sigma$, steadily increases.
Consequently, the opening angle of a jet increases as it propagates through the plasma~\cite{Chesler:2015nqz}.
The energy lost from the jet to the plasma corresponds to energy density along the string  traveling along blue geodesics falling
into the horizon.
Clearly, geodesics with smaller  $\sigma$, i.e.~with smaller initial angle, propagate the farther before reaching the horizon. 
The thermalization length of the jet, $x_{\rm therm}$, is the distance that the endpoint travels before it reaches the horizon.
It is apparent from the figure that jets with a smaller $\sigma_0$, meaning a narrower initial opening angle, lose their energy more slowly and have a longer $x_{\rm therm}$~\cite{Chesler:2015nqz}.
Figure adapted from Ref.~\cite{Chesler:2015nqz}.
}
\end{figure}

With the goal of making this paper more self-contained, we shall spend the remainder of this
introduction reviewing some aspects of various previous holographic
calculations that provide the basis and context for how we (and others) use selected strings in a holographic gauge theory as models for jets.  Along the way we shall also set up key elements of
our analysis that follows.
Fig.~\ref{fig:NullString}, adapted from Ref.~\cite{Chesler:2015nqz}, provides a good starting point as it illustrates many key features of how to build a holographic model for jets in plasma.
The depth of the black
hole horizon in the AdS direction is $1/(\pi T)$; it sets the inverse 
temperature of the strongly coupled
plasma in the
field theory.
If the Figure had been drawn in vacuum, there would be no horizon and the blue geodesics would all be straight lines.  The blue geodesics in the Figure as drawn curve downward because of the
presence of the horizon, which is to say because of the presence of the plasma.
The essence of the holographic dictionary is that a depth $z$ into the bulk 
corresponds to a length-scale $z$ in the gauge theory.  Hence, if a (red) bit of energy
propagates along a (blue) trajectory that is a straight line heading rightward and downward with
some angle $\sigma$ in the Figure, this bit of energy is holographically dual to energy in the gauge theory
that expands in size linearly in time as it propagates rightward.  This is to say it is dual to a flux of energy in the gauge theory that fills a cone with an opening angle proportional to $\sigma$.
In order to model a jet in vacuum, with some unchanging value of its jet mass,
we must therefore find a string whose endpoint travels with some constant 
downward angle $\sigma_0$.  That is, in vacuum the downward angle of the string endpoint, $\sigma_0$, in the gravitational description is proportional to the opening angle of the
``jet'' in the ${\cal N}=4$ SYM theory that we use to model a jet with that ratio of jet mass to jet energy in QCD.
With this correspondence established, we can now read many qualitative features of
jet quenching in a strongly coupled gauge theory directly from the Figure. The fact that string energy travelling along blue geodesics falls into the black hole is the 
gravitational description
is energy lost by the jet as the jet excites a wake in the plasma.
The fact that the string endpoint trajectory, like any of the blue trajectories, curves downward
corresponds to the fact that the opening angle of the jet expands as the jet propagates through
the plasma, losing energy.
The fact that trajectories whose initial downward angle is smaller go farther before falling into
the horizon corresponds to the fact that jets whose initial opening angle, proportional to $\sigma_0$,
is smaller lose energy more slowly, over a longer distance, and  travel farther through the plasma before thermalizing in the plasma.

Via the calculation that corresponds 
to the blue geodesics in  Fig.~\ref{fig:NullString} falling into the horizon, the authors of Refs.~\cite{Chesler:2014jva,Chesler:2015nqz} 
obtained an analytic expression for the rate at which these jets, or better to say
these models for jets constructed in strongly coupled ${\cal N}=4$ SYM theory, lose energy as they propagate through the strongly coupled
plasma:
\begin{equation}
\label{eq:energylossresult0}
\frac{1}{E_{\rm init}} \frac{dE_{\rm jet}}{dx} = - \frac{4 x^2}{\pi x_{\rm therm}^2 \sqrt{x_{\rm therm}^2 - x^2}}.
\end{equation}
where $E_{\rm init}$ is the initial energy of the jet represented by the null string 
in Fig.~\ref{fig:NullString} and where $x_{\rm therm}$ is the thermalization distance of the jet, namely the distance
that the string endpoint travels before falling into the horizon.  
This distance is related to the initial downward angle of the string endpoint, $\sigma_0$, by~\cite{Chesler:2015nqz}
\begin{equation}
T x_{\rm therm} = \frac{\Gamma\left(\frac{1}{4}\right)^2}{4\pi^{3/2}}\frac{1}{\sqrt{\sigma_0}}\ ,
\label{eq:xtherm-sigma0}
\end{equation}
which quantifies the fact that jets whose opening angle is initially smaller travel farther through the plasma.
For a jet that travels a small distance and loses only a small fraction of its initial energy, 
we can expand and integrate (\ref{eq:energylossresult0}), obtaining~\cite{Chesler:2015nqz}
\begin{equation}
\frac{dE_{{\rm jet}}}{dx}=-E_{{\rm jet}}\frac{256\pi^{7/2}}{\Gamma\left(\frac{1}{4}\right)^{6}}T^{3}x^{2}\sigma_{0}^{3/2}+\mathcal{O}(x^{4})\ .
\end{equation}
When we construct our ensemble of strings with which we shall model an ensemble of jets
in plasma in Section~\ref{sec:model}, we shall follow Ref.~\cite{Rajagopal:2016uip} in treating the proportionality constant
between the initial downward angle of the endpoint of a string, $\sigma_0$, and the opening
angle of the jet that we wish to model with that string as a free parameter. This is the first of two free parameters in our model; we shall denote it by $a$ and will define it precisely in Section~\ref{sec:model}. 

It is  important to realize that not all string configurations that can be constructed
in the gravitational dual of ${\cal N}=4$ SYM theory behave like the string in Fig.~\ref{fig:NullString}. Not all by any means.
Strings have a nonzero tension, but in Fig.~\ref{fig:NullString} the string tension 
does not affect the dynamics of the red string because each bit of string is following a null 
geodesic.  
One can (and in fact we will in the next Section) 
instead construct string configurations with different initial
conditions in which the string worldsheet is, at least initially, not null and 
in which the string tension affects
the dynamics of the string to such a degree that the downward angle at which
the endpoint of the string moves changes substantially as it propagates {\it even in vacuum.}  
It is not immediately apparent how such a string can serve as a model for a jet, since on the face
of it it would seem to correspond to a jet whose virtuality changes substantially after the jet has been created, something that does not happen in QCD
since a high energy jet once formed interacts at most softly with 
other jets or with the underlying event, meaning that by momentum conservation the virtuality of a jet in QCD hardly changes.
In the gravitational description of strongly coupled ${\cal N}=4$ SYM theory, though, there are (non-null) strings in vacuum in which an end of the string 
feels a force from the rest of the string that changes its trajectory. 
This means that in strongly coupled
${\cal N} = 4$ SYM theory there are configurations in which a flux of energy behaves completely differently from a jet in QCD. (This is unsurprising: most string configurations in the gravitational
dual do not correspond to anything that looks like a jet.  And, furthermore, in strongly coupled 
${\cal N}=4$ SYM theory hard processes do not produce jets~\cite{Hatta:2008tx,Hofman:2008ar}.)
It remains an open question whether a subset of strings whose end point trajectories change
their downward angle can nevertheless be used directly as models for jets; investigating this would require computing the gauge theory energy flux and looking for instances where virtuality and opening angle do not change, even when the string end point trajectory does. In the present work, we follow a more straightforward approach.
We choose
to model jets in QCD by choosing strings in the gravitational description
of strongly coupled ${\cal N}=4$ SYM theory whose endpoints follow a trajectory
with some constant downward angle $\sigma_0$ in vacuum, curving downward only
because of the presence of the black hole horizon, modeling jets whose opening
angles change only because they are propagating through plasma.
The simplest way that we know of choosing strings that constitute good models
for jets in QCD is to choose null strings, as in Fig.~\ref{fig:NullString}, following an approach that goes back to Ref.~\cite{Chesler:2008wd}.

Our discussion to this point has left the distribution of energy along the string unspecified.
For null strings in vacuum, whatever distribution of energy along the string we choose
initially (respecting open string boundary conditions) will simply propagate unchanged along the
blue null geodesics (which are straight in vacuum).  Since a null string like that in 
Fig.~\ref{fig:NullString} propagates for an initial period of time $\ll 1/(\pi T)$ as if it were in vacuum, 
we have considerable freedom in choosing the initial energy density along the string.
After the string has propagated through the plasma for a distance that is $\gg 1/(\pi T)$,
its shape is no longer semicircular, as in vacuum.  In fact, after propagation through 
the plasma it takes on the 
shape of a segment of the string that describes an infinitely heavy quark being
dragged through the plasma~\cite{Chesler:2015nqz}, a shape that was
first worked out in Refs.~\cite{Herzog:2006gh,CasalderreySolana:2006rq,Gubser:2006bz}.
As the string propagates through the plasma over a distance $\gg 1/(\pi T)$ and blue trajectory after blue trajectory peels away and falls into the horizon, eventually the only 
aspect of the initial distribution of energy along the string that matters is the
energy that is initially very close to the endpoint of the string.
For this reason, the authors of Ref.~\cite{Chesler:2015nqz} chose an initial distribution of energy
along the string that takes the near-endpoint form $\propto 1/(\sigma^2\sqrt{\sigma-\sigma_0})$ dictated by the open string boundary conditions
everywhere along the string.
Although operationally reasonable, the logic behind this choice
is not fully satisfactory since it is based upon using the form of the string energy density after the string has propagated for
a long distance through the plasma to choose the distribution of energy along the string
initially, when the string is still behaving as if it were in vacuum.
It would be better to have an argument based upon the physics of strings in vacuum
for choosing the initial distribution of energy along the string.
We shall remedy this lacuna in Section~\ref{sec:strings}.

In Section~\ref{sec:strings} we study the dynamics
of strings that are initially {\it not} null.
As anticipated, in vacuum their endpoints do not follow trajectories 
with a constant downward angle, meaning that they (initially)
represent objects whose virtuality is not obviously conserved
which makes it unclear how they can be used to model jets in QCD.
As an extreme example, extreme in the sense that they are the least
apparently jet-like of any of the strings we analyze, 
we include strings
similar to the ones considered in
Ref.~\cite{Ficnar:2013wba} in which the downward angle of the string endpoint changes suddenly.
From 
our analysis of their dynamics, however, we find that
a large class of strings that are initially not null
become null strings as they fall in the AdS vacuum: after a 
certain ``nullification time'' that we compute, every bit of string moves
along a null geodesic, as anticipated in Ref.~\cite{Chesler:2014jva,Chesler:2015nqz}.
After nullification, the string endpoints follow
trajectories with a constant downward angle meaning that after nullification
all these strings end up becoming jet-like.
And, quite remarkably, we find that for a rather 
diverse set of initial conditions for the energy density along
the string, as long as we don't make the string null initially (in which case
the energy distribution would not change in vacuum) after the string nullifies 
the distribution of energy density along
the string has evolved such that it is approximated 
by a scaling form parametrized only by the downward
angle of the string endpoint after nullification.  Near the string endpoint this scaling
form agrees with the expression for the distribution of energy density along
the string obtained from the near-endpoint expansion of Refs.~\cite{Chesler:2014jva,Chesler:2015nqz}, as it must.
We show that in plasma, which is to say when the gravitational 
description of the physics includes a horizon, the strings
that we analyze nullify while they are still far above the horizon
meaning that their nullification occurs as it would in vacuum.

The results for the dynamics of strings that are initially not null, in Section~\ref{sec:strings},
motivate our construction, in Section~\ref{sec:model}, of an ensemble of
null strings as a model for an ensemble of jets.  We choose the
distribution of energy along a null string in this ensemble with a specified
initial downward angle $\sigma_0$ according to the scaling form
 for this distribution, namely the scaling form attained by 
 initially non-null strings.  This means that we are using
 nontrivial string dynamics in (the dual of) strongly coupled ${\cal N}=4$ SYM theory, dynamics
 that does not itself appear to be jet-like, to determine how to distribute the energy density along the strings
 in the ensemble of null
 strings that we  subsequently use to model an ensemble of jets.
 Another way of describing this is that among all the possible null strings that
 can be constructed in the dual of ${\cal N}=4$ SYM theory, the subset that we
 choose to use in our ensemble of strings are those
 with a particular scaling form for their energy distribution such that they
 can be formed
 either by starting with null strings from the beginning or by starting with
 strings that are initially not null and evolving them until they nullify.

 As in Ref.~\cite{Rajagopal:2016uip}, we choose the
 distribution of the initial jet energies and opening angles for the jets
 in our ensemble from perturbative QCD
 calculations as appropriate for
 QCD jets in proton-proton collisions.  For each jet, we choose the initial distribution
 of energy along the string that represents that jet in our model according to the scaling
 form obtained via our holographic analysis of the nullification of strings.
 This specifies the jet shape for each jet in our ensemble.
 Remarkably given that the strongly coupled 
 dynamics by which the strings nullify has no apparent analogue in QCD, 
 we find in Section~\ref{sec:results} that, upon fitting the single parameter $a$,
 our model yields a very good description of the mean jet shape in QCD, as measured
 in proton-proton collisions by the CMS collaboration~\cite{Chatrchyan:2013kwa}.

%

With our ensemble of strings fully specified, including via using the distribution
of energy along each string taken from the scaling form obtained via our holographic
analysis of nullification, in Section~\ref{sec:results} 
we send the ensemble through an expanding cooling
droplet of hydrodynamic fluid (described in Section~\ref{sec:model}).
In so doing, we introduce a second free parameter in the model, which we denote by $b$, which is the proportionality constant between the temperature of the QCD
plasma that we are modeling and the temperature of the ${\cal N}=4$ SYM 
plasma (with more degrees of freedom) that we are using as a model.
We choose $b$ such that the modification in the number of jets with a given energy
in the ensemble  after it has passed through the droplet of plasma relative to that  in
the initial ensemble is comparable to that seen in data.
We then compute the modification to the mean jet shape, and compare to experimental
data~\cite{Chatrchyan:2013kwa}.
We find a narrowing in the jet shape at small angles that is comparable to that seen
in data, but because we are not including the contribution to reconstructed jets coming
from the wake in the plasma our model cannot describe the modification to the mean jet shape at larger angles.
We then consider an ensemble of dijets, with the dijet asymmetry distribution chosen
to reproduce that measured in proton-proton collisions, and compute the
modification to this distribution caused by passage through the droplet of plasma.
Here again we compare to experimental data from heavy ion collisions 
at the LHC~\cite{Chatrchyan:2012nia}.
We close with a look ahead at possible future improvements to the model.
Our results for the modification of the dijet asymmetry distribution are promising but they are not in quantitative agreement with the data; this motivates a future analysis of
an ensemble of trijets, since in reality (and unlike in our model dijet ensemble) much of the 
dijet asymmetry seen in proton-proton collisions comes from events in which there is a third jet present.

\section{String Dynamics}

\label{sec:strings}

With the aim of studying the dynamics and evolution of an ensemble
of null strings in $\mathcal{{N}}=4$ SYM plasma as a model for
an ensemble of jets, and in particular for the purpose of
choosing the shape of the distribution of energy density along the null strings,
we begin with a study of
strings in 4+1-dimensional AdS space that are, initially, not null. In holography, a pair
of light quarks is represented by an open fundamental string in AdS
\cite{Karch:2002sh}. The 5-dimensional metric in AdS which corresponds
to a constant-temperature plasma in the 4-dimensional $\mathcal{N}=4$
SYM theory on its boundary is 
\begin{equation}
\diff s^{2}=\frac{L^{2}}{z^{2}}\left(-f(z)\diff t^{2}+\diff\vec{x}_{\perp}^{\:2}+\diff y^{2}+\frac{\diff z^{2}}{f(z)}\right)\,,\label{metric-1}
\end{equation}
where $z$ is the additional direction in AdS space, $f(z)=1-z^{4}/z_{h}^{4}$,
and the black hole is located at $z=z_{h}\equiv 1/\pi T$. Here $\vec{x}_{\perp}$
and $y$ are field theory coordinates specifying the transverse plane
and the beam direction, respectively. This metric is an exact solution
to Einstein's equations for a constant-temperature plasma.  We shall later
(in Section~\ref{sec:model}) choose a temperature $T$  that varies in space and time
so as to model an expanding cooling droplet of plasma but for
a spatially-varying temperature profile this model neglects transverse
flow, fluid viscosity, and gradients.

The dynamics of strings in this geometry are most conveniently
solved numerically using the Polyakov action (see Refs.~\cite{Chesler:2008wd,Chesler:2008uy,Morad:2014xla}):
\begin{equation}
S_{P}=-\frac{T_{0}}{2}\int d\tau_{ws}d\sigma_{ws}\sqrt{-\eta}\,\eta^{ab}\,\partial_{a}X^{\mu}\partial_{b}X^{\nu}\,G_{\mu\nu},
\end{equation}
with $T_{0}=\sqrt{\lambda}/2\pi$ the string tension with $\lambda$
the 't Hooft coupling, $\tau_{ws},$ $\sigma_{ws}$ the string worldsheet
coordinates, $G_{\mu\nu}$ the bulk AdS metric and $\eta_{ab}$ the
string worldsheet metric, which determines the gauge choice in mapping
$\tau_{ws}$ and $\sigma_{ws}$ coordinates to target space coordinates
$X^{\mu}$. The gauge degree of freedom $\eta_{ab}$ can be solved
for by varying the action, giving the constraint equation:
\begin{equation}
\gamma_{ab}=\frac{1}{2}\eta_{ab}\eta^{cd}\gamma_{cd},
\end{equation}
where $\gamma_{ab}=\partial_{a}X\cdot\partial_{b}X$ is the induced
metric on the string worldsheet. Since the worldsheet metric is a
gauge choice, the functions $X^{\mu}(\tau_{ws},\,\sigma_{ws})$ can
be chosen to make the numerics more straightforward. Since we will
typically solve the equations of motions in steps along $\tau_{ws}$,
this for instance requires that different parts of the string cover
the part of spacetime in a similar pace in the $\tau_{ws}$ variable.
This can be done by defining \cite{Chesler:2008uy}
\begin{equation}
\eta_{ab}\equiv\left(\begin{array}{cc}
-\Sigma(X^{\mu}) & 0\\
0 & 1/\Sigma(X^{\mu})
\end{array}\right),
\end{equation}
with the stretching function $\Sigma(X^{\mu})$ , which is commonly
chosen to cancel singularities in the equations of motion. In all the string evolutions 
presented in this paper, we shall use 
\begin{equation}
\Sigma=\left(\frac{1-z}{1-z_{0}}\right)^{\alpha}\left(\text{\ensuremath{\frac{z_{0}}{z}}}\right)^\beta
\end{equation}
with $\alpha$ and $\beta$ typically 1 or 2.
From the action we can also obtain the target space energy-momentum
density:
\begin{equation}
\pi_{\mu}^{a}(\tau_{ws},\,\sigma_{ws})=\frac{1}{\sqrt{-\eta\,}}\,\frac{\delta S_{{\rm P}}}{\delta(\partial_{a}X^{\mu}(\tau_{ws},\sigma_{ws}))}=-T_{0}\,\eta^{ab}\,\partial_{b}X^{\nu}\,G_{\mu\nu}.
\end{equation}
The energy-momentum density at some time $t$ is then given by
\begin{equation}
p_{\mu}(t,\,\sigma_{ws})=\frac{\sqrt{\lambda}}{2\pi}\sqrt{-\eta}\left(\pi_{\mu}^{\tau}(t,\,\sigma_{ws})-\pi_{\mu}^{\sigma}(t,\,\sigma_{ws})\frac{\partial_{\sigma_{ws}}t}{\partial_{\tau_{ws}}t}\right)\ .\label{eq:energy}
\end{equation}
We shall analyze the string dynamics throughout
in classical Einstein gravity and using classical equations of motion,
which means everything is done in the limit of strong coupling.
In the strict $\lambda\rightarrow\infty$ limit there are no quasi-particles
and hence also no distinguishable quark-antiquark pairs. In AdS this
is dual to the statement that in this strict limit 
creating a string requires an infinite energy of
order $\mathcal{O}(\sqrt{\lambda})$. Our limit hence has to interpreted
as a large but finite coupling, where it does make sense to consider
a quark-antiquark pair, dual to a string. Later, when we quote numerical
results we shall always take $\lambda=5.5$, as in Ref.~\cite{Gubser:2006qh}.

We create each string at a single point $(z_0,t_0,x_0)$ in the AdS spacetime.
Without loss of generality we can set $x_0=0$ and $t_0=0$.  One way of varying
the initial conditions for our strings is to vary $z_0$.  We must also specify
initial conditions for the velocity of the string in the AdS spacetime
as a function of the string worldsheet parameter $\sigma_{ws}$, subject
to open string boundary conditions. 
The aim of this Section is to introduce several classes of initial conditions
for the string and to show that when we choose the string to
not be null initially its dynamics turn it into a null string 
(a string where each segment travels along an independent null geodesic)
after a period of time, which we compute.
We find it striking that, although the nullification process occurs
through strongly-coupled dynamics which may have no direct analog
in the dynamics of jets in QCD,
it yields an approximate scaling form for the distribution of energy
density along the string after the string nullifies.  We shall use this scaling
form in Sections \ref{sec:model} and
\ref{sec:results} when we follow the evolution of an ensemble of null strings,
which serve as models for jets,
as they pass through an expanding and cooling droplet of plasma.

\begin{figure}
\begin{centering}
\includegraphics[width=13cm]{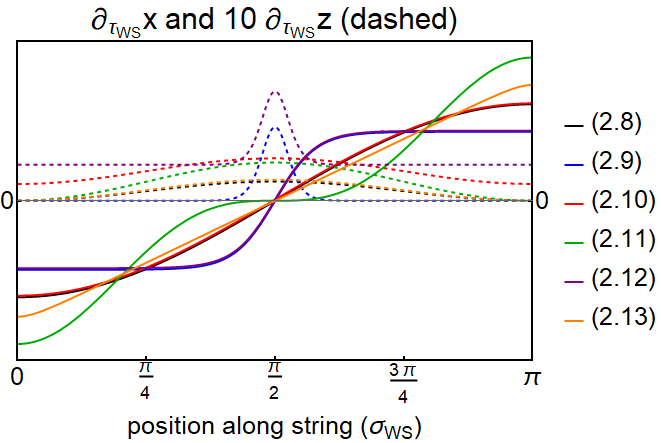}
\par\end{centering}
\caption{
We show the shapes of the  profiles that we use for the initial conditions on the velocities in AdS of
the strings that we analyze,
as given in Eqs.~(\ref{eq:firstIC}-\ref{eq:lastIC}) and Eqs.~(\ref{eq:firstTau}-\ref{eq:secondTau}). 
The solid
curves show the initial velocity of the string in the $x$-direction 
as a function of the worldsheet coordinate $\sigma_{ws}$.
(The string endpoints are at $\sigma_{ws}=0$ and $\pi$; the string midpoints are at $\sigma_{ws}=\pi/2$.)
The dashed curves give the initial velocity in
the $z$-direction (``downward'' into the bulk of AdS), enhanced by a factor of 10
for visibility. As discussed in the text, the small
but nonzero $z$ velocity is useful numerically to make $\partial_{\tau_{ws}}t$
continuous at the midpoint of the string. 
The relationship between the worldsheet time coordinate
$\tau_{ws}$ and the AdS time $t$ is given in terms of these functions
in Eq.~(\ref{eq:tau_and_t}).}
\label{fig:SixInitialConditions}
\end{figure}

We shall analyze six classes of initial conditions for the strings, depicted
in Fig.~\ref{fig:SixInitialConditions}.
The 
initial velocity of the string in the $x$-direction is given as a function of $\sigma_{ws}$ by
the six expressions:
\begin{eqnarray}
\partial_{\tau_{ws}}x(\sigma_{ws}) & = & A\,\cos(\sigma_{ws}),\,\label{eq:firstIC}\\
\partial_{\tau_{ws}}x(\sigma_{ws}) & = & A\left(\frac{1}{2}\tanh\left(4\left(\sigma_{ws}-\frac{\pi}{2}\right)\right)  - \text{sech}^{2}(2\pi)\sin(2\sigma_{ws})\right)\label{eq:secondIC}\\
\partial_{\tau_{ws}}x(\sigma_{ws}) & = & A\,\cos(\sigma_{ws}),\,\label{eq:angleIC}\\
\partial_{\tau_{ws}}x(\sigma_{ws}) & = & A\,\cos^{3}(\sigma_{ws}),\,\label{eq:ini4}\\
\partial_{\tau_{ws}}x(\sigma_{ws}) & = & A\left(\frac{1}{2}\tanh\left(4\left(\sigma_{ws}-\frac{\pi}{2}\right)\right)  -\text{sech}^{2}(2\pi)\sin(2\sigma_{ws})\right)\label{eq:angleIC2}\\
\partial_{\tau_{ws}}x(\sigma_{ws}) & = & A\left(\sigma_{ws}-\frac{1}{10}e^{-10(\pi-\sigma_{ws})}+\frac{e^{-10\sigma_{ws}}}{10}-\frac{\pi}{2}\right)\,\label{eq:lastIC}
\end{eqnarray}
with $A$ a parameter that we specify as described  below. The velocity in the holographic
 $z$-direction
is given by 
\begin{equation}
\partial_{\tau_{ws}}z(\sigma_{ws})=\frac{A}{200}\Bigl(1-\cos(2\sigma_{ws})\Bigr) + A\sigma_{s} 
\label{eq:firstTau}
\end{equation}
for (\ref{eq:firstIC}), (\ref{eq:angleIC}), (\ref{eq:ini4}) and (\ref{eq:lastIC})
and by
\begin{equation}
\partial_{\tau_{ws}}z(\sigma_{ws})=\frac{A}{40}\left[\tanh\left(10\Bigl(\sigma_{ws}-\frac{\pi}{2}\Bigr)+\frac{1}{2}\right)+1\right]\left[\tanh\left(10\Bigl(\frac{\pi}{2}-\sigma_{ws}\Bigr)+\frac{1}{2}\right)+1\right]+A\sigma_{s}
\label{eq:secondTau}
\end{equation}
for (\ref{eq:secondIC}) and (\ref{eq:angleIC2}).  
This $z$-velocity is small but useful to ensure that
$\partial_{\tau_{ws}}t$ is continuous at $\sigma_{ws}=\pi/2$. 
We shall set the parameter $\sigma_s=0$ in (\ref{eq:firstIC}), (\ref{eq:secondIC}), (\ref{eq:ini4}) and (\ref{eq:lastIC}).
This means that in these four classes of initial conditions, the string initially has a nonzero
velocity in the $z$-direction only near $\sigma_{ws}=\pi/2$.  
And, in these four classes
of initial conditions the initial velocity of the string endpoints (at $\sigma_{ws}=0$ and $\pi$) 
are {\it horizontal}, corresponding initially to a collimated flow of energy with vanishing opening angle.
If it were possible to create a jet with zero initial virtuality in QCD, 
its virtuality would remain zero; it would never fragment into a shower and would never fill a cone.
No production mechanism for a collimated object like this is known in QCD.
We shall see below, however, that in ${\cal N}=4$ SYM the strongly coupled dynamics
ensures that an object created with zero opening angle like this does
not stay that way.  
The strongly coupled dynamics turns this initially collimated 
object into something that later becomes jet-like.  Which is to say that the string (which
is initially not null) nullifies.
In initial conditions (\ref{eq:angleIC}) and (\ref{eq:angleIC2}) we choose a nonzero value of the parameter $\sigma_{s}$ such that the initial downward
angle of the string endpoints is $0.7^\circ$ and $2.8^\circ$ respectively. 

To complete the specification of the initial conditions that we shall analyze, we note that the
velocity in the time direction follows from the constraint equation
$\eta_{00}=0$ and from our assumption that the string starts at a single point.
It is given by
\begin{equation}
\partial_{\tau_{ws}}t=\sqrt{(\partial_{\tau_{ws}}z(\sigma_{ws}))^{2}+(\partial_{\tau_{ws}}x(\sigma_{ws}))^{2}}.\label{eq:tau_and_t}
\end{equation}
Last, we describe how we choose the value of the parameter $A$.
By conformal invariance we can keep one parameter fixed and then vary
other parameters without loss of generality. 
In this Section, we choose to fix $A$ such that the energy (\ref{eq:energy})
is $E=1000$. We performed several numerical checks, verifying
for all of our evolutions that the constraint equations are satisfied
and that the total energy as obtained from (\ref{eq:energy}) is conserved
up to our numerical precision ($10^{-4}$ or better).

\subsection{The transition to null strings}

\label{sec:nullstrings}

At the start of the evolution of the strings with initial conditions
(\ref{eq:firstIC}), (\ref{eq:secondIC}), (\ref{eq:ini4}) and (\ref{eq:lastIC}),
in which the initial downward angle of the string endpoints vanishes,
the string tension is a crucial ingredient in determining the dynamics
of the string.  The string stretches initially, losing kinetic energy as it does so.
This effect is especially strong
near the boundary of the AdS spacetime, which is to say near the endpoint of the string,
where the larger proper distance
due to the large AdS metric factor requires the string to have a large
initial energy to off-set the potential energy cost of the stretching.
This suggests that there should be a sense in which strings 
of this type which start closer to the AdS boundary (smaller $z_0$) 
have larger energy.   We shall make this precise below.

We shall see that what happens to these strings
is that the string tension succeeds in pulling the string endpoints away from
the AdS boundary and giving them a nonzero downward angle even though
their initial downward angle vanishes.   (This dynamics has no apparent analogue in the physics
of jets in QCD.)
After some time of evolution, then, the strings have fallen into
the bulk AdS space and the increasing kinetic energy becomes dominant
over the potential energy. 
At this stage the string tension 
no longer has a significant effect on the dynamics of the string.
To a good approximation, each segment of the string
travels on a null geodesic: the string nullifies.
Focusing on the endpoint of the string, the string tension initially curves
its trajectory downward, away from the AdS boundary, but after some
time its downward angle stops increasing and it henceforth follows a null geodesic
with a constant downward angle.
We shall call the downward angle reached by the string endpoint as the
string becomes null $\sigma_0$.  We can now state the precise sense in
which strings which start closer to the AdS boundary have larger energy: we 
shall show that, for strings
whose initial downward angle vanishes, the smaller the $z_0$ we choose
the larger the  energy we must choose if we wish to end up with
a specified value of $\sigma_0$ after nullification.   

If instead of starting off with a string whose endpoint initially moves horizontally
we start off with a string that is {\it initially} null, the discussion above changes
completely.  There is no stretching effect. The string tension never plays a significant
role in the string dynamics. And, the initial downward angle of the string endpoint keeps
its initial nonzero value $\sigma_0$ throughout.  These are the strings that we shall use as models of QCD jets in vacuum.  However, in this case the dynamics leaves
the distribution of energy density along the string unchanged also, which gives us no
clue as to what distribution to choose.  
The initial conditions (\ref{eq:angleIC}) and (\ref{eq:angleIC2}) with initial downward
angles for the string endpoints that are nonzero
are not null, but
they (in particular (\ref{eq:angleIC2})) are much closer to being null than any of
our initial conditions for strings whose endpoints are initially horizontal.
We therefore expect that these strings should nullify more quickly, with less rearrangement
of the energy density along the string. And, we expect that for these
strings it need not be the case that reducing $z_0$ means increasing the energy
needed to achieve a specified $\sigma_0$.

Note that
since in vacuum null geodesics
in AdS are just straight lines it is clear that a nullified string
is fully specified in vacuum 
by how much energy goes downward at what angle in AdS.
As we illustrated in Fig.~\ref{fig:NullString}, it is also true
in plasma
that a nullified string is completely specified
by the initial downward angle of each bit of energy along the string.
However, in this case the blue null geodesics in Fig.~\ref{fig:NullString} 
curve downward toward the black hole horizon meaning that the downward angle
of each bit of energy along the string increases.

\begin{figure}
\begin{centering}
\includegraphics[width=12.5cm]{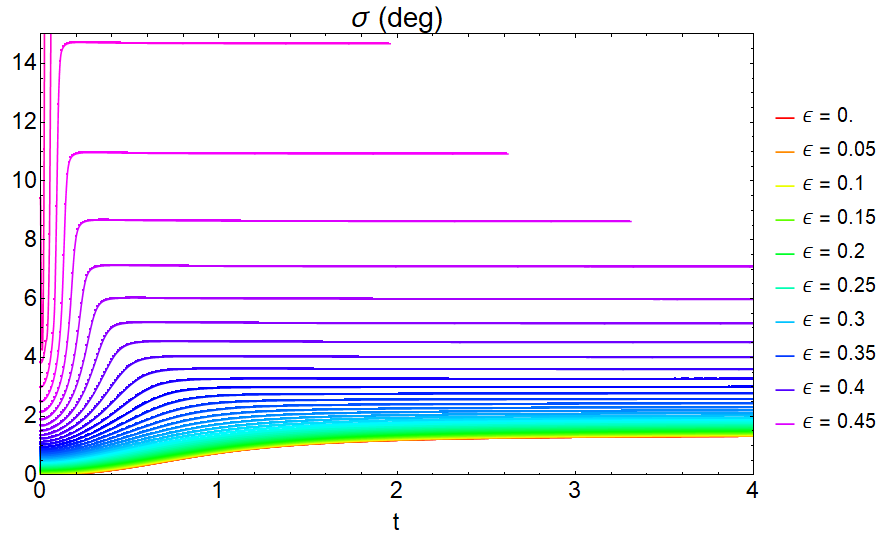}
\includegraphics[width=12.5cm]{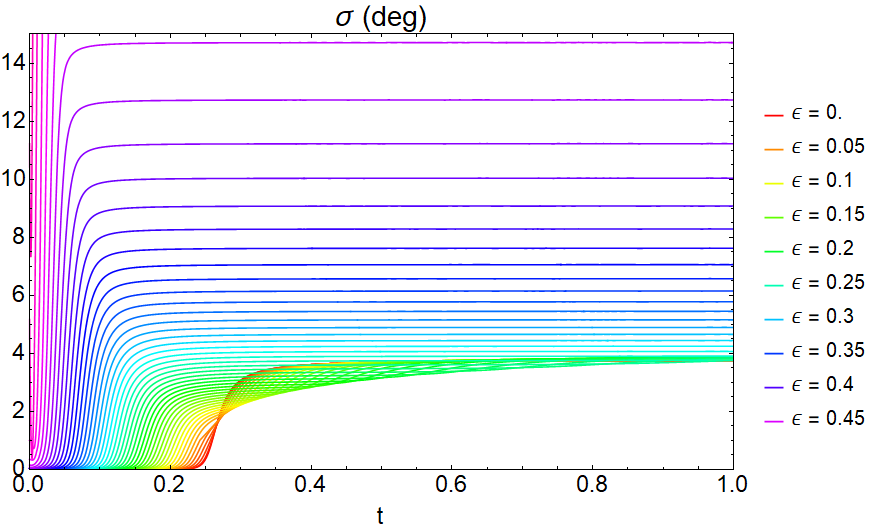}
\par\end{centering}
\caption{We depict the time evolution in vacuum of two 
strings whose endpoints
initially move horizontally, with zero downward angle.
The upper panel is for the initial condition (\ref{eq:firstIC}) with $z_0=0.02$ and
the lower panel is for the initial condition (\ref{eq:secondIC}) with $z_0=0.01$.
Each curve shows the time evolution of the downward angle $\sigma$ (not to be confused
with the worldsheet coordinate $\sigma_{ws}$) of the point on the 
string such that a fraction $\epsilon$ of the total energy of the string
is found between that point and the string endpoint.
Hence, $\epsilon=0$ corresponds to the string endpoint and $\epsilon=0.5$ corresponds
to the string midpoint.
The calculation is done entirely in vacuum. And, in the AdS vacuum null geodesics
are trajectories with a constant downward angle $\sigma$.
We see that each point on the string ``nullifies'': after some time, each bit of energy along the string moves along a null geodesic with some constant $\sigma$.
After nullification, the string endpoints move along trajectories
with nonzero downward angles.
\label{fig:two-examples}}
\end{figure}

Figure \ref{fig:two-examples} shows two examples of 
string
evolution in vacuum, for a string with the initial condition (\ref{eq:firstIC}) created at
$z_{0}=0.02$ (above) and for a string with the initial condition (\ref{eq:secondIC}) created at 
$z_{0}=0.01$ (below).
We plot the time evolution of the downward angle of the point on the string above
which 
a fraction $\epsilon$ of the string energy is found. That is, $\epsilon$ starts
at 0 at the string endpoint and $\epsilon=0.5$ corresponds to the string
midpoint, which by symmetry moves straight downward into the AdS bulk. 
For both strings in the figure, the initial downward angle of the string endpoint is zero.
As a consequence of the string tension, during the early time dynamics
each segment of the string changes its downward angle $\sigma$.
For each segment of the string, though, after some time passes
its downward angle no longer changes. That is, the string nullifies.
The constant nonzero downward angle of the trajectory that the
 string endpoint follows after nullification, $\sigma_0$, is
 $1.8^\circ$ in the upper panel of Fig.~\ref{fig:two-examples} 
 and $3.8^\circ$ in the lower panel.

 The early time dynamics is particularly dramatic in the
 lower panel of Fig.~\ref{fig:two-examples}, where the endpoint of the string propagates almost horizontally
 for a time before relatively suddenly turning downwards with a nonzero angle
 that soon becomes constant.  
 These strings, with the initial condition (\ref{eq:secondIC}), 
describe a flow of energy that is initially collimated, with zero opening angle, 
before
 later, suddenly, acquiring a substantial opening angle.
 We shall describe this dynamics further
 in subsection~\ref{sec:EndpointStrings}.
First, though, we shall provide a further description of the nullification process
in vacuum and shall then illustrate that it works quite similarly in plasma --- because 
nullification happens while the strings are still far enough from the horizon that
their dynamics is similar to that in vacuum.  We shall see that even though the strings
with initial conditions (\ref{eq:secondIC}) have dramatic dynamics early
on, after they nullify and become jet-like they look rather similar to
the jet-like strings that form starting from the other initial conditions that we analyze.

\subsubsection{Nullification in vacuum}
\label{sec:nullification-vacuum}

\begin{figure}
\begin{centering}
\includegraphics[width=12cm]{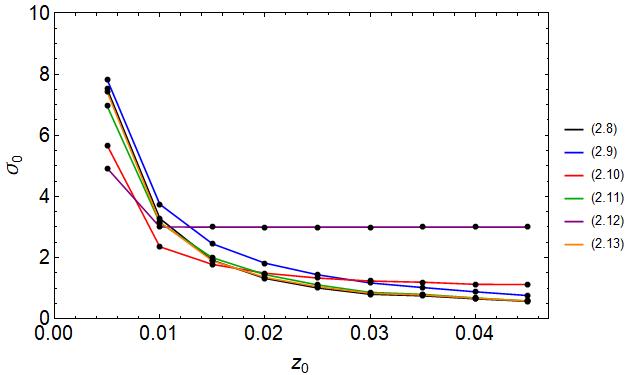}
\includegraphics[width=12cm]{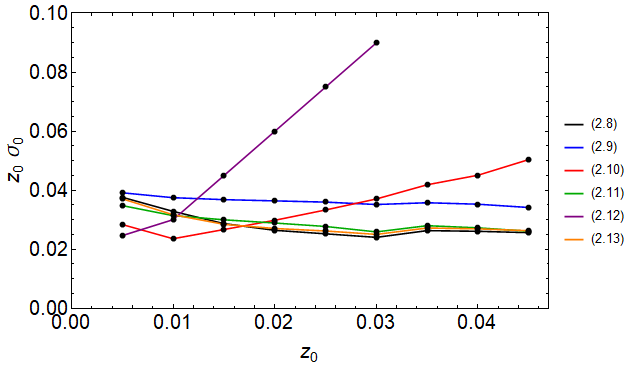}
\par\end{centering}
\caption{
For each of the six classes of initial conditions
in Eqs.~(\ref{eq:firstIC}-\ref{eq:lastIC}),
we plot the downward angle $\sigma_0$ reached by the endpoint of
the string after it nullifies, which is to say after the downward angle no longer changes.
In the upper panel, we show how $\sigma_0$ varies as we change the AdS 
depth $z_0$ at which we initialize the string, while keeping the energy of the
string fixed.
We find that for strings with the initial conditions
(\ref{eq:firstIC}), (\ref{eq:secondIC}), (\ref{eq:ini4}) and (\ref{eq:lastIC})
in which the downward angle of the string endpoint is initially zero, 
to a good approximation the opening angle $\sigma_0$ reached
by the string endpoint after nullification is given by
$\sigma_{0}\sim\frac{1}{E\,z_{0}}$, with a prefactor $c$ that depends weakly
on the initial string profile.  
For the initial conditions (\ref{eq:angleIC}) and (\ref{eq:angleIC2}) 
in which the downward angle of the string endpoint is initially
nonzero and in which the strings are closer to null from the beginning,
this relationship is not satisfied.
\label{fig:sigma0}}
\end{figure}

\begin{figure}
\begin{centering}
\includegraphics[width=12.5cm]{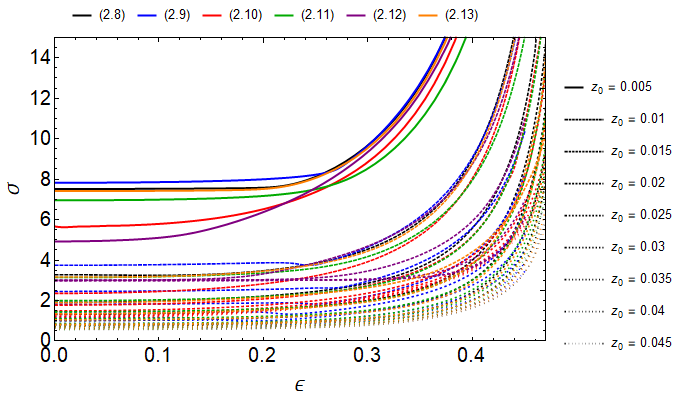}
\includegraphics[width=13.cm]{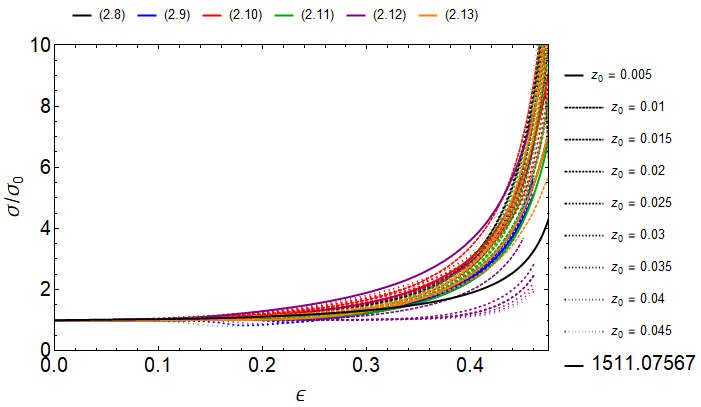}
\par\end{centering}
\caption{After nullification, the point on the string above which a fraction $\epsilon$
of the energy of the string is found moves along a null geodesic with some
constant downward angle $\sigma$.  In the upper panel, we plot this $\sigma$
as a function of $\epsilon$, after 
nullification, for strings with each of our six classes of initial conditions with
varying values of $z_0$.  $\sigma_0$  is the value of $\sigma$ at $\epsilon=0$.
In the lower panel, we rescale each of the many curves in the upper panel by its own $\sigma_0$.
We find that for the strings with the initial conditions
(\ref{eq:firstIC}), (\ref{eq:secondIC}), (\ref{eq:ini4}) and (\ref{eq:lastIC})
in which the initial downward angle of the string endpoint was zero, after nullification
the rescaled curves in the lower panel all take on a rather similar shape.
This means that, for these initial conditions, the nullification dynamics rearranges the way that energy
density is distributed along the string as a function of $\sigma/\sigma_0$ such that it reaches an approximate scaling form.
The purple curves show that for initial conditions like (\ref{eq:angleIC2}) that are close to null from the beginning, because the energy density distribution along the string 
hardly changes it need not reach the scaling form.
\label{fig:Full-string-results} }
\end{figure}

\begin{figure}
\begin{centering}
\includegraphics[width=13cm]{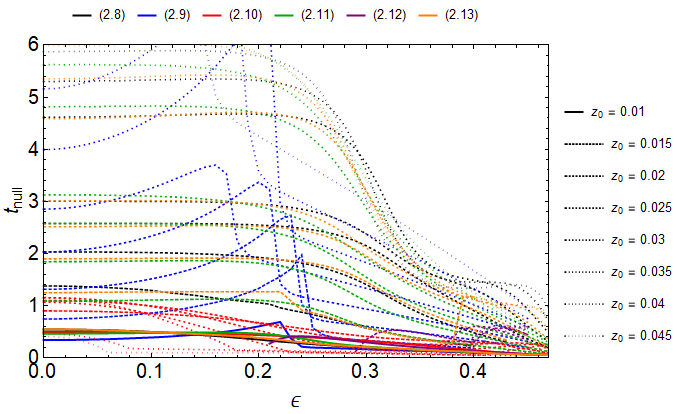}
\includegraphics[width=13cm]{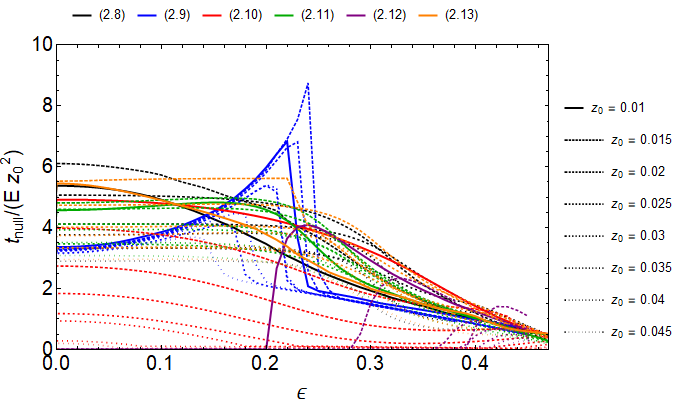}
\par\end{centering}
\caption{For a point on the string above which
a fraction $\epsilon$ of the string energy is found, we define the 
nullification time $t_{\rm null}$ as the time that the angle found
in Fig. \ref{fig:Full-string-results} is within 10\% of its final
value. In the upper panel we plot $t_{\rm null}$ as a function of $\epsilon$
for the strings that we have analyzed.
In the lower panel, we plot $t_{\rm null}/(E z_0^2)$, 
and find that the nullification time is roughly proportional to $z_0^2$ 
for the strings whose endpoints had an initial downward angle of zero.
Those strings which we initialized with initial conditions (\ref{eq:angleIC}) with their endpoints moving with a nonzero downward angle
are closer to null from the beginning, meaning that it is no surprise that they
nullify faster. 
\label{fig:nullification-times}}
\end{figure}

In this subsection, we shall compute 
the nullification timescales and
the resulting distribution of energy along the string 
as a function of the downward angle of the trajectory followed by a bit of string
after nullification, for strings whose initial conditions are given
by Eqs.~(\ref{eq:firstIC}-\ref{eq:lastIC})
that nullify in vacuum. 
The nullification process depends on the initial velocity profile,
the energy of the half-string $E$ and the 
AdS depth $z_0$ of the point at which we initialize the string.
We have analyzed nullification for the six classes of initial conditions
presented above with varying values of $z_0$. 
Without loss of generality, in vacuum we can choose units of energy such that
$E=1000$. 
For each string that we evolve, we compute the opening angle $\sigma_0$
with which the string endpoint is descending into the bulk after the string
has nullified 
as well as the
time $t_{\text{null}}$ it takes for the endpoint to reach this angle
within accuracy of 10\%. In Figures \ref{fig:sigma0}, \ref{fig:Full-string-results} and \ref{fig:nullification-times} 
we show these observables
for our strings, and more.

In Figure \ref{fig:sigma0} (bottom) we see that 
if we choose initial conditions from one of our classes of initial conditions
in which the initial downward angle of the string endpoint is zero then,
after the initial phase of the string dynamics, when the string nullifies 
its endpoint is moving downward into the AdS bulk at a constant
angle $\sigma_0$ that is well approximated by
\begin{equation}
\sigma_{0}\sim\frac{c}{E\,z_{0}},
\end{equation}
for a profile-dependent constant $c$.  Here, we have reinstated the
$E$-dependence by dimensional analysis. 
(Strings that are initially closer to null, like those with initial conditions
(\ref{eq:angleIC}) and in particular (\ref{eq:angleIC2}), do not satisfy
this relationship.)
In Ref.~\cite{Chesler:2015nqz}, it was shown analytically
that in a limit in which $\sigma_0\rightarrow 0$ 
null strings as in Fig.~\ref{fig:NullString} have an
energy $E\propto \sigma_0^{-3/2}$.  We now see that
we can reproduce this limit by choosing a sequence of strings with zero
initial downward angle and increasing $E$ 
as long as we initialize the strings at a $z_0$ that we choose
to be $ \propto E^{-1/3}$.  The strings in this sequence will
nullify with a $\sigma_0\propto E^{-2/3}$.

In Fig.~\ref{fig:Full-string-results} we plot the downward angle $\sigma$
of every point on each of the strings that we have studied, after nullification, rather than just focusing
on the string endpoint.  
We find that for the strings with initial conditions
in which the initial downward angle of the string endpoint was zero, after nullification
the downward angle of a bit of string, scaled by the downward angle $\sigma_0$ of the
endpoint of that string, takes on an approximate
scaling form as a function of what fraction of the energy is found above that bit of string.
This is equivalent to saying that the energy distribution along
the string takes on an approximate scaling form after nullification.
In Section~\ref{sec:model} when we use an ensemble of null strings
to model an ensemble of QCD jets, we shall choose to distribute the 
energy density along these null strings according to the scaling form
found in the lower panel of Fig.~\ref{fig:Full-string-results}. Specifically, 
we shall take  the form for $\sigma/\sigma_0$ as a function of $\epsilon$
found after nullification for strings that start out at $z_0=0.005$ with the initial condition (\ref{eq:firstIC}).
This scaling form is the principal result of this Section; it is the result from this Section that
we shall employ when we model an ensemble of jets in Section~\ref{sec:model}.

In the lower panel of Fig.~\ref{fig:Full-string-results} the scaling form that 
we find after nullification is compared
to the 
result from Ref.~\cite{Chesler:2014jva}
(the black curve labelled 1511.07567),
in which the near-endpoint approximation for the energy density distribution
as a function of $\sigma$, namely
$e(\sigma)\propto\frac{{1}}{\sigma^{2}\sqrt{{\sigma-\sigma_{0}}}}$,
was employed for the entire string.
This provides a good approximation to the scaling form that we have found
for $\epsilon \lesssim 1/4$, which is to say for the half of the energy of the string
that is closer to its endpoint.  Farther away from the endpoint, the near-endpoint
approximation does not describe the scaling
form that we have found.



Finally, in Figure~\ref{fig:nullification-times}
we show the nullification times for 
every point on each of the strings that we have studied.
There it can be seen that for those strings which
start out far from null, with their endpoints moving horizontally,
and subsequently nullify via the strongly coupled dynamics
that we have focused on in this Section do so after a
nullification time that is around $(2-6) E z_0^2$,
depending somewhat on the initial condition as well as on the position on the string.


\subsubsection{Nullification in plasma}

We shall now analyze the dynamics of strings with the initial conditions
that we have introduced above in plasma, rather than in vacuum.
When the strings are initially far above the black hole horizon, near the boundary,
they are in a region of the spacetime where the metric is nearly the
same as in vacuum.  This means that  strings which nullify
quickly compared to the time it takes them to fall close to the black hole horizon
will nullify via dynamics that is nearly the same as the dynamics in vacuum
that we have analyzed above.
We shall show in this subsection that, to good accuracy, this is indeed the case for strings
with the initial conditions that we have chosen.

\begin{figure}
\begin{centering}
\includegraphics[width=12cm]{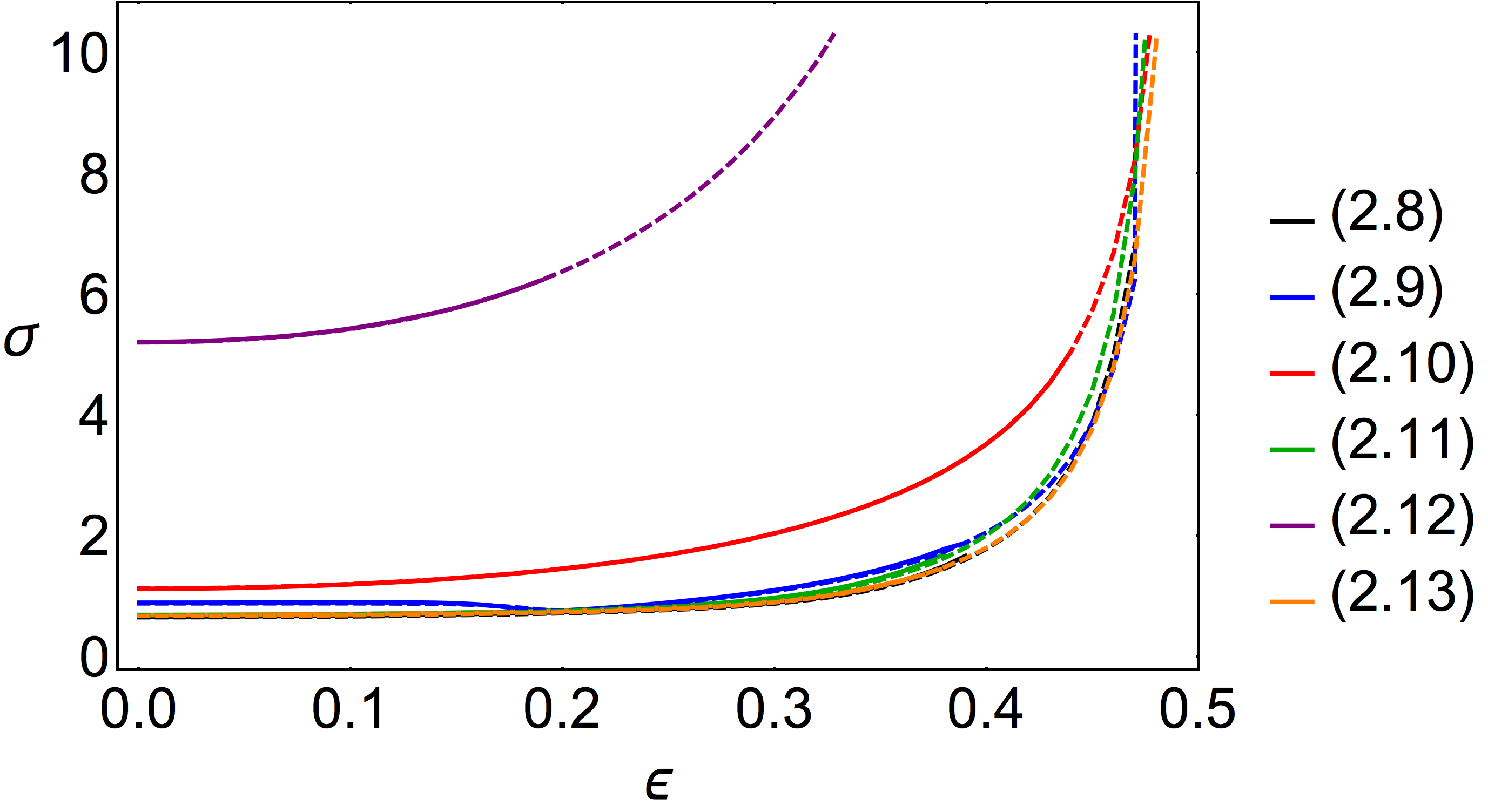}
\
\
\includegraphics[width=12cm]{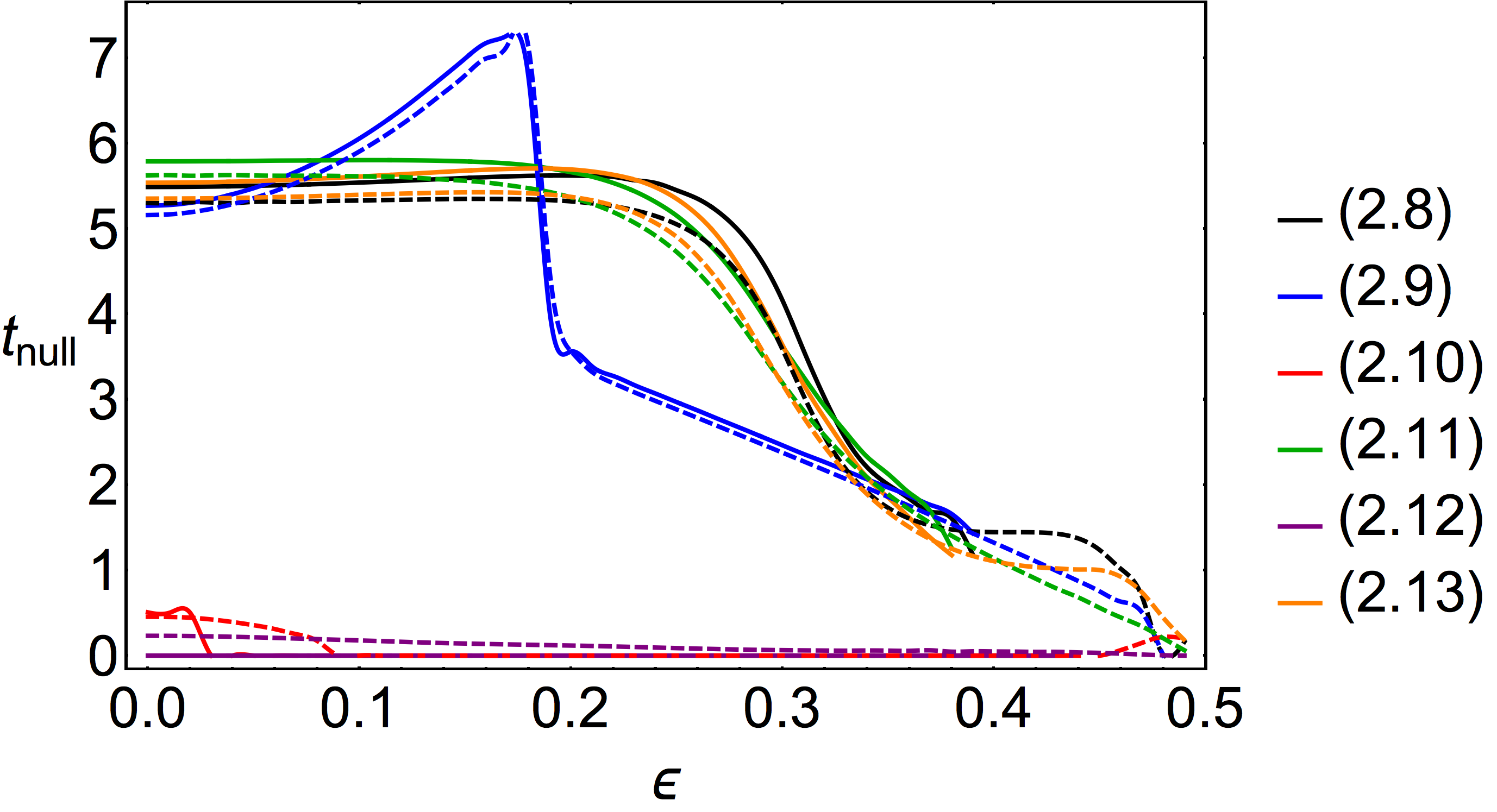}
\par\end{centering}
\caption{We compare the downward angles of the string endpoints after nullification $\sigma$ defined as described in the text (top panel) and the nullification times (bottom panel)
for strings that are produced and then nullify in plasma (solid curves) to those that
are produced and then nullify in vacuum (dashed curves) as in the previous subsection.
All strings were produced at $z_0=0.04$ except for those with initial conditions (\ref{eq:angleIC2}), which
were produced at $z_0=0.005$.
\label{fig:vacuum-vs-plasma}}
\end{figure}

In vacuum, null geodesics are straight lines. This simplifies many
computational aspects of studying nullification
in vacuum compared to in plasma, because the angles of null geodesics
stay fixed and the deviation from null is given by the deviation of
the trajectory from a straight line and hence is easily assessed and quantified. 
As we discussed in Section~\ref{sec:intro}, in the AdS black hole spacetime dual
to the plasma, null geodesics curve downward toward the horizon and a string gets represented
by a congruence of null geodesics that
loses energy to a wake in the plasma and ultimately thermalizes over a distance
$x_{\rm therm}$, the distance that the string endpoint travels before falling into the horizon.
Because nullification happens
relatively quickly compared to the thermalization time  $x_{\rm therm}$, we expect that it
happens near the boundary where the spacetime is close to vacuum AdS. 
However, to test this we must have a way of  assessing whether the
string has nullified that can be applied in the regime where null geodesics
curve downwards, namely in the regime that is not near the boundary.
At any given time $t$ we construct a null geodesic that is tangent to the
actual trajectory of the string endpoint (or for that matter to any chosen bit of string) 
and follow that null geodesic backwards/upwards
all the way to the boundary.   In the near-boundary region, this null geodesic
is straight and has some downward angle. Once the string has nullified,
its endpoint is following a null geodesic.  This means that once
the string has nullified, if we go to a later time and repeat the exercise of shooting
a null geodesic that is tangent to the string endpoint trajectory back upwards we
will find the same null geodesic with the same initial downward angle as we found
at the earlier time.  We define the angle $\sigma_0$ as this initial downward angle, defined
from the near-boundary slope of the 
null geodesic that the string endpoint follows at late times.  And we define the nullification
time $t_{\rm null}$ as the time after which this initial downward angle changes by less than 10\%.
Defined in this way, both $\sigma_0$ and $t_{\rm null}$ can be compared
directly to their values in vacuum, allowing for the quantitative comparison
between the nullification dynamics in plasma to that in vacuum
depicted in Figure \ref{fig:vacuum-vs-plasma}. 
We see that the nullification dynamics is similar indeed, confirming that nullification
happens quickly enough that, in plasma, it happens in the near-boundary regime where
the spacetime is very similar to vacuum AdS.

We conclude that the strings that we investigate nullify
far above the horizon, near the boundary.  This means that the expression (\ref{eq:xtherm-sigma0})
provides a reasonable approximation to the relationship between the distance that they
travel between nullification and thermalization,
$x_{\rm therm}$, and their initial downward angle after 
nullification $\sigma_0$, defined as described above. 
It also completes the justification for how we shall model jets
in  Section~\ref{sec:model}: we shall use null strings as in Fig.~\ref{fig:NullString}
with a specified initial downward angle $\sigma_0$ and with energy density
distributed along the string according to the scaling form that we have
found via our analysis in subsection~\ref{sec:nullification-vacuum} of the dynamics of how strings nullify in vacuum.

\subsection{Strings with most of their energy at their endpoint}
\label{sec:EndpointStrings}


Before turning to modeling jets, we close this section with a further look
at the nullification of strings with 
the initial condition (\ref{eq:secondIC}), noting that this is close to a particular
string initial condition from Ref.~\cite{Ficnar:2013wba} 
in which all of the energy and momentum of
the string is initially localized at its endpoint, and hence travels
initially in the same direction.
We have seen that the dynamics of these strings before nullification is dramatic.
However, these strings nullify like the others and have the benefit that they
allow a partial
analytic treatment, one that will allow us to confirm (in this special case) some of the scaling
expressions that we found above more generally, but numerically.
It has been shown in the literature that 
if we choose initial conditions in which all of the string energy is localized at its endpoint
with the string endpoint initially moving
parallel to the boundary at some $z_0\ll 1/(\pi T)$, with
zero downward angle, then
the endpoint loses energy
according to \cite{Ficnar:2013wba,Ficnar:2013qxa}
\begin{equation}
\frac{dE}{dx}=-\frac{\sqrt{\lambda}}{2\pi}\frac{1}{z_0^2}\ .
\end{equation}
%
%
The endpoint 
will follow a straight line (i.e. a null geodesic) at constant $z_0$ 
until all its energy
has been depleted, which happens after a distance:
\begin{equation}
x_{{\rm snap}}=\frac{2\pi}{\sqrt{\lambda}}E\,z_{0}^{2}.
\label{eq:snapback}
\end{equation}
At that point the endpoint cannot continue further along its initial null geodesic and
must change direction (a ``snapback''), as can clearly be seen in the lower panel of 
Fig.~\ref{fig:two-examples}. 
(With $E=1000$, $\lambda=5.5$ and $z_0=0.01$, the expression (\ref{eq:snapback}) yields $x_{\rm snap}=0.27$, in agreement with the behavior seen in that Figure.)
This means that  at this point in time
the string is clearly different from a null string and hence we find
that for this particular type of strings  $t_{{\rm null}}\geq (2\pi/\sqrt{\lambda})E\,z_{0}^{2}$, consistent
with the approximate scaling that we found more generally, but numerically, above.

After this snapback, nullification occurs: the string endpoint finds itself moving
along a new null geodesic with some nonzero downward angle $\sigma_0$ 
that does not change further. 
%
%
In order for the endpoint to continue on a null geodesic without another
snapback we find a condition on the initial endpoint energy $E_{0}$,
its starting position $z_0$, and $\sigma_0$, the opening angle after nullification, that is given by
\begin{equation}
E_{0}\geq\int_{z_{0}}^{\infty}dz\,\frac{dE}{dz}
=-\int_{z_0}^\infty dz \frac{\sqrt{\lambda}}{2\pi}\frac{1}{z^{2}\sqrt{1-f/R^{2}}}
\end{equation}
where $R=-f\sqrt{\dot{x}^{2}+\dot{z}^{2}}/\dot{x}=1/\cos(\sigma_0)$, which is to say by
\begin{equation}
E_0\geq
\int_{z_{0}}^{\infty}dz\,\frac{1}{z^2 \sin(\sigma_0)}=\frac{1}{\sin(\sigma_{0})z_{0}}\ ,
\end{equation}
which is again consistent with the more general scaling that we found numerically above.


\section{A Model for an Ensemble of Jets in Heavy Ion Collisions}

\label{sec:model}

We are now ready to construct the ensemble of null strings
in the dual description of ${\cal N}=4$ SYM theory that
we shall use as a model for an ensemble of jets in heavy ion collisions.
The first step is to understand the relationship between
the energy density distributed along an individual null string
and the shape of the individual jet that this string represents.
Here by shape we mean the distribution of $P_{\rm out}$, the outward-directed
flux of power at infinity, as a function of the angle $r$ measured from the center of the jet.
We use the result
from Ref.~\cite{Chesler:2014jva}: 
%
\begin{equation}
\frac{dP_{\rm out}}{d\cos r}=\frac{1}{2}\int_{\sigma_0} d\sigma\frac{e(\sigma)}{\gamma(\sigma)^{2}[1-v(\sigma)\cos r]^{3}}\,,\label{power_out}
\end{equation}
where as in Fig.~\ref{fig:NullString} we have parametrized the null string worldsheet by $\sigma$, the initial downward angle of a blue null geodesic along which a bit of energy travels, where
$e(\sigma)$ is the energy density along the string as a function of $\sigma$ with
 $E_{\rm final}=\int_{\sigma_0} d\sigma e(\sigma)$, where
$\gamma(\sigma)\equiv (1-v(\sigma)^{2})^{-1/2}$, and where $v(\sigma)=\cos\sigma$
for a null geodesic. This formula is relevant for a null string propagating through
a finite droplet of QGP that emerges from that droplet and then propagates onward to infinity in vacuum.
The domain of integration is over the angles $\sigma$ that label those blue null geodesics
that do not fall into the black hole.
We will choose an ensemble of strings
with differing values of the initial downward angle of the string
endpoint, $\sigma_0$. We shall specify the probability distribution for $\sigma_0$
and for the initial energy of the string below.
For an individual string in the ensemble with some particular value of $\sigma_0$,
we choose $e(\sigma)$ to be given by the approximate scaling form that we found
in Section~\ref{sec:nullification-vacuum} for the distribution
of energy along the string after nullification for strings 
that were initially not null, whose endpoints initially had no downward angle.
The approximate scaling form found after nullification is illustrated in the lower panel of 
Fig.~\ref{fig:Full-string-results}, where what is plotted is $\sigma/\sigma_0$ 
as a function of $\epsilon$, for points on the string above which a fraction $\epsilon$ of 
the total energy of the string.  Specifically, we use the curve from Fig.~\ref{fig:Full-string-results}
for strings that start out at $z_0=0.005$ with the initial condition (\ref{eq:firstIC}).
For a string of total energy $E$ whose
downward endpoint angle at nullification is given by $\sigma_{0}$, 
we denote the curve in the
lower panel of Fig.~\ref{fig:Full-string-results} by $\sigma(\epsilon)$
and express the jet shape $P_{out}(r)$
as a function of the angular coordinate $r$ away from the jet axis as 
\begin{equation}
P_{out}(r)=\int d\epsilon \frac{E \sin r}{\gamma(\sigma(\epsilon))^{2}[1-v(\sigma(\epsilon))\cos r]^{3}}\ .
\label{power_out_2}
\end{equation}
This relation allows us to compute the shape of individual model
jets in the $\mathcal{{N}}=4$ SYM gauge theory 
corresponding to null strings with a specified initial downward angle $\sigma_0$.

Next, we must specify the distribution of the initial energy $E$ and downward
angle $\sigma_0$ of the strings  in the ensemble  that we shall use to model
an ensemble of jets.
We shall then be able to compute the mean jet shape in the ensemble in vacuum.
We will then construct an ensemble of dijets, using the probability
distribution of the dijet asymmetry measured in proton-proton collisions.
We will also need a model for the evolution of
the plasma and a distribution for the starting points and directions
of the jets in the transverse plane.
After sending our jets through an expanding cooling droplet of 
strongly coupled  ${\cal N}=4$ SYM plasma, we shall investigate
how the mean jet shape and the distribution of the dijet asymmetry
are modified by passage through the plasma.

As in Ref.~\cite{Rajagopal:2016uip}, we shall utilize perturbative
QCD calculations of the probability distribution 
for a
useful measure of the opening angle of a jet in QCD
defined by
\begin{equation}
C_{1}^{(1)}\equiv \sum_{i,j}z_{i}z_{j}\frac{\theta_{ij}}{R}\,,\label{eqn:c11-1}
\end{equation}
where the sum is over all pairs of hadrons in the jet, $\theta_{ij}$
is the angle between hadrons $i$ and $j$, and $z_{i}$ is the momentum
fraction of hadron $i$. We shall consider jets recontructed with the anti-$k_t$
algorithm~\cite{Cacciari:2008gp} with reconstruction parameter $R=0.3$, as in the CMS data that
we shall compare our results to below.
We shall (quite arbitrarily) take the quark
and gluon fractions each to be 0.5 in the formulae for the $C_{1}^{(1)}$
distribution calculated
in perturbative QCD in Ref.~\cite{Larkoski:2014wba}. The opening angle
of a holographic jet is proportional
to the downward angle of the string endpoint $\sigma_0$~\cite{Chesler:2015nqz},
but we have no direct analogue 
of $C_{1}^{(1)}$ since the holographic calculation does not have
hadrons meaning that we cannot calculate Eq.~(\ref{eqn:c11-1}) explicitly.
Therefore, as in Ref.~\cite{Rajagopal:2016uip}  we shall
take 
\begin{equation}
C_{1}^{(1)}=a\sigma_{0}\ ,
\end{equation}
introducing a free parameter $a$ in our model.
This allows us to translate the perturbative QCD calculations 
for the distribution of $C_{1}^{(1)}$ into a probability distribution
for the initial downward angle $\sigma_0$ of the strings
in our ensemble.  Note that this probability distribution depends on the 
initial jet energy $E$. We complete the specification of our ensemble
of strings by
choosing a distribution of initial jet energies which falls
as $E^{-6}$. 

\begin{figure}
\begin{centering}
\includegraphics[width=12cm]{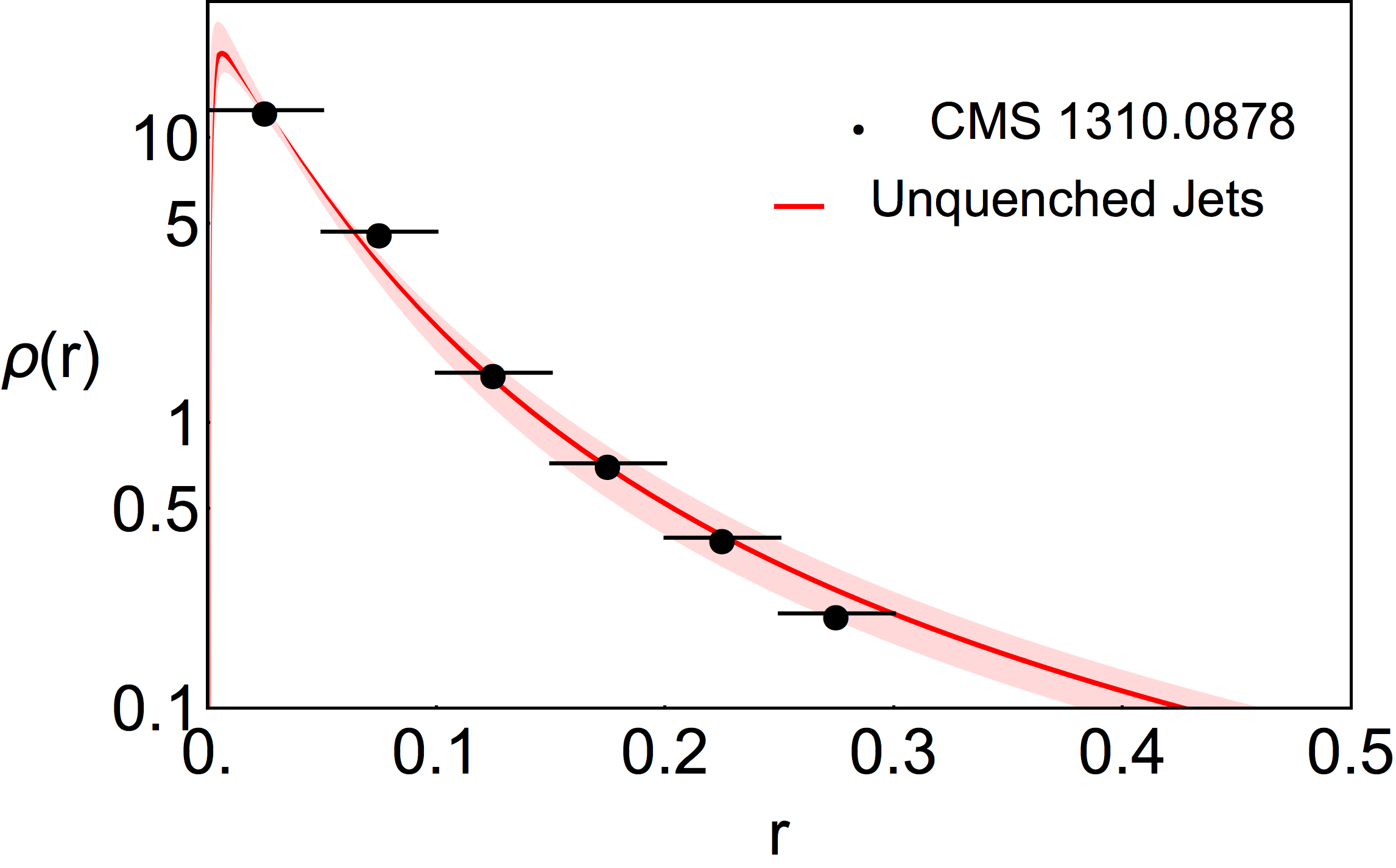}
\par\end{centering}
\caption{Mean jet shape in vacuum as a function of $r$, the
angle in $(\eta,\phi)$-space from the center of the jet, 
computed from the ensemble of null strings described in the text compared
to CMS measurements of the mean jet shape for
jets with energy above 100 GeV reconstructed with anti-$k_t$ reconstruction
parameter $R=0.3$  in  proton--proton collisions with 
$\sqrt{s_{\rm NN}}=2.76$~TeV at the LHC
from Ref~\cite{Chatrchyan:2013kwa}, shown as black symbols.   
The pink band shows the results obtained from our 
ensemble of strings for a range in the free parameter given by $a=1.8-2.5$, with
$a=2$ shown in dark red. \label{fig:rhoplot} 
Although the effects of doing so
are negligible here, the experimental
data and the results of our calculations have been smeared in order 
to take into account the CMS jet energy resolution, as described 
in Ref.~\cite{Chatrchyan:2013kwa} and later in the text.
}
\end{figure}

The differential jet shape for an individual jet is the power $P_{out}(r)$
as a function of the angle $r$ from the jet axis, as given 
in Eq.~(\ref{power_out_2}).
The  (normalized) meanx
jet shape is the average of the individual jet shapes over the ensemble.
We plot our results for the mean jet shape in Fig.~\ref{fig:rhoplot},
binning them in bins of width $\Delta r=0.05$ for consistency
with the CMS data, which we also show in the Figure. 
We find that upon making a suitable choice for the free parameter in the model $a$
we obtain a rather good description of the mean jet shape measured in proton-proton
collisions!
The result shown in Figure~\ref{fig:rhoplot} has the
best fit value $a=2$ shown in red, which is in reasonable agreement with
the crude estimate of $a\sim1.7$ given in \cite{Rajagopal:2016uip}
for smooth jets, as well as a band of predictions corresponding to varying $a$
from 1.8 to 2.5. 
We find it pleasing, and perhaps even remarkable, that
even though we picked the initial energy distribution along the null
strings that we are using to model jets from a strongly coupled
calculation of the dynamics of strings as they nullify that is quite different from
the dynamics of jets in QCD, the mean jet shape that we obtain
agrees so nicely with measurements made in proton-proton collisions. 

To compute the modification to the mean jet shape in our ensemble
caused by the passage of the jets through the strongly coupled plasma,
we need a model for the dynamics of the droplet of plasma.
We assume boost invariant longitudinal expansion and  initialize the droplet of
plasma at a proper time $\tau=1$~fm/c after the collision.
As in Ref~\cite{Rajagopal:2016uip}, we make the overly simplified
assumption that our null strings are produced at the same time
that the hydrodynamic plasma is initialized, thus completely neglecting the possibility
of energy loss before the plasma hydrodynamizes.  Again as in Ref.~\cite{Rajagopal:2016uip},
we also make the 
overly simplified assumption that all quenching stops and the strings propagate as in vacuum
after the droplet of plasma has cooled below $T=175$~MeV.
The droplet of $\mathcal{{N}}=4$ SYM plasma and its evolution are encoded, via
the AdS-CFT correspondence, in changes to the 5-dimensional metric
in AdS space. An expanding and cooling droplet of plasma in the field
theory corresponds to a black hole in 5-dimensional AdS space whose horizon
is expanding  in the spatial directions while shrinking ``downward'', away from the AdS boundary, 
in the $z$-direction.
As in Ref.~\cite{Rajagopal:2016uip}, we take a simple blast-wave profile to model the temperature evolution
in the transverse plane and assume boost-invariant longitudinal expansion, choosing
\begin{equation}
T(\tau,\vec{x}_{\perp})=b\left[\frac{dN_{{\rm ch}}}{dy}\frac{1}{N_{{\rm part}}}\frac{\rho_{\text{part}}(\vec{x}_{\perp}/r_{\text{bl}}(\tau))}{\tau\,r_{\text{bl}}(\tau)^{2}}\right]^{1/3}.
\end{equation}
Here, $\tau\equiv\sqrt{t^{2}-z^{2}}$ is the proper time, $\rho_{\text{part}}(\vec{x}_{\perp})$
is the participant density in the transverse plane as given by an optical Glauber model, 
and
$r_{\text{bl}}(\tau)\equiv\sqrt{1+(v_{T}\tau/R)^{2}}$ with $v_{T}=0.6$
and $R=6.7$~fm. We consider 2.76 TeV Pb-Pb collisions
at mid-rapidity and 0-10\% centrality at the LHC, and based upon averaging the results for 0-5\% and 5-10\% centrality from Ref.~\cite{Abbas:2013bpa} we take the number of participants to
be $N_{\text{part}}\simeq 356$ and based upon summing
the results for pions, kaons and protons in Ref.~\cite{Abelev:2013vea} we take $dN_{{\rm ch}}/dy\simeq 1599$.
$b$ is the second free parameter in our model; we shall use it to parameterize
differences in the number of degrees of freedom between $\mathcal{{N}}=4$
SYM and QCD, meaning that we should model a QCD plasma with temperature $T$
by an ${\cal N}=4$ SYM plasma with some lower temperature. 
The constant $b$ is a measure of the multiplicity per entropy, 
and for a QCD plasma is $b\approx 0.78$~\cite{Rajagopal:2016uip}. Since we are modeling this plasma
with an ${\cal N}=4$ SYM plasma, in our model we must scale the temperature
that we use down by choosing a value
of $b$ that is substantially smaller than this~\cite{Rajagopal:2016uip}.

In
this work, we consider an ensemble of $\approx150,000$ jets which
sample distributions in jet opening angle, energy, and the starting
position and direction of the jets within the droplet of plasma. 
We take the initial position of the quark-antiquark pair in the transverse
plane to be distributed according to a binary scaling distribution
proportional to $\rho_{\rm part}(\vec{x}_{\perp})^{2}$ with their directions randomly distributed
in the transverse plane.  We have described our choice for the initial
distribution of the energy and opening angle, and hence downward angle
for the null string endpoint,  above.
For the analysis of the dijet asymmetry, we additionally sample the initial dijet asymmetry distribution, which increases the size of our ensemble by a factor of roughly 30.

\begin{figure}
\begin{centering}
\includegraphics[width=12cm]{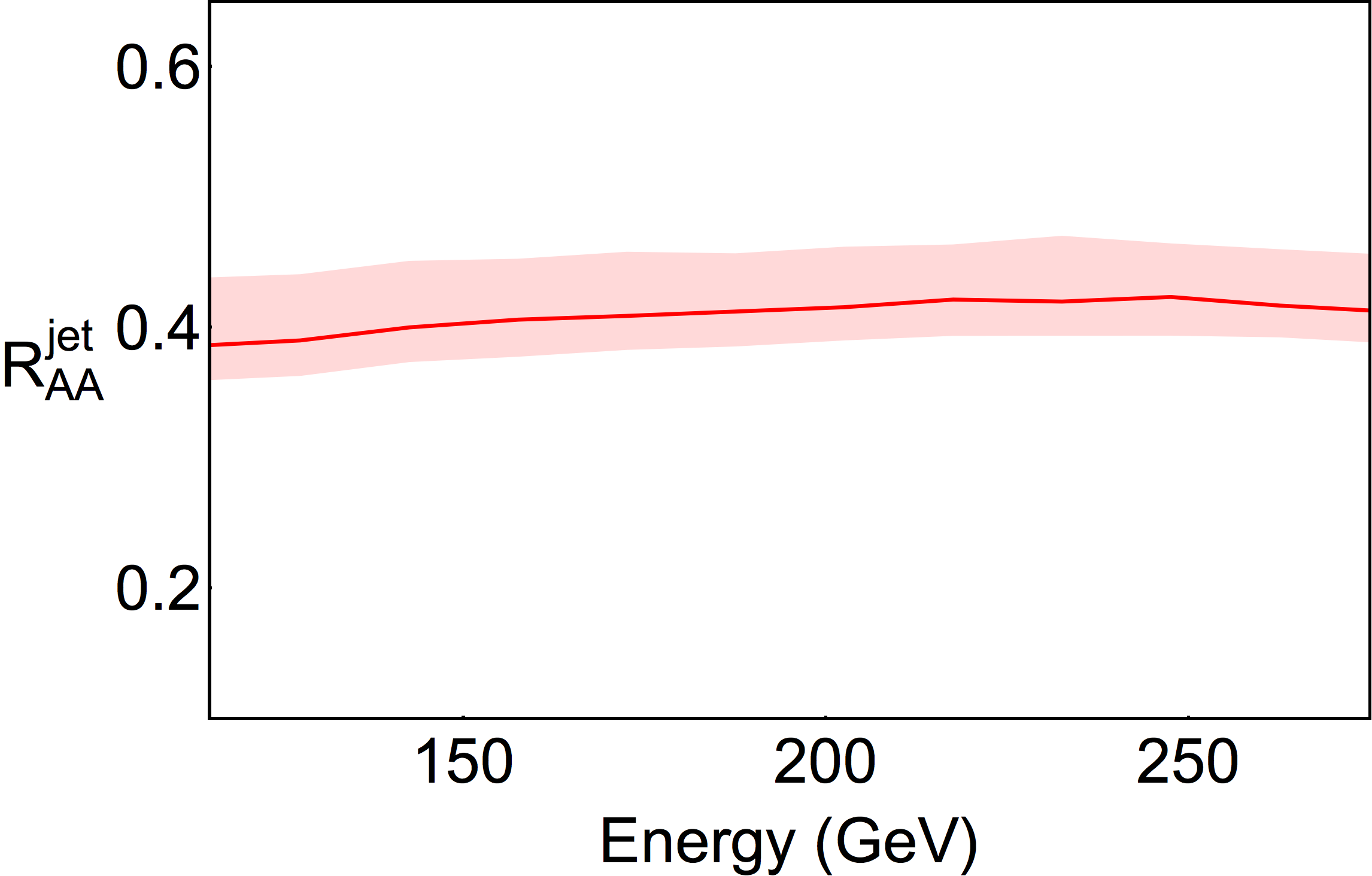}
\par\end{centering}
\caption{
Jet suppression $\RAA^{\rm jet}$, the ratio of the number of 
jets with a given energy in our ensemble after the jets have been quenched
via passage through the expanding cooling droplet of plasma to the number of jets with the same energy in the initial ensemble, before quenching. 
As in Fig.~\ref{fig:rhoplot}, the pink band
shows the results obtained from our ensemble of strings for  a range in the parameter $a$ given by $a=1.8-2.5$, with $a=2$ shown in dark red.
We have chosen the value of $b$,  the second free parameter in the model, defined in the text, to be $b=0.21$ so as to obtain a jet suppression that is comparable to 
experimental measurements of jet $\RAA$, 
for example those in Refs.~\cite{Raajet:HIN,Aad:2014bxa,Khachatryan:2016jfl}.
  \label{fig:RAAplot} }
\end{figure}

After we send our ensemble of strings through the expanding cooling droplet of
plasma, we recompute the energies and opening angles (given by $a$ times
the downward angle of the string endpoint after it emerges from the droplet of plasma)
of each string in the ensemble.  In the next Section, we shall compute the modification to 
various 
observables from the ensemble after passage through the plasma.
The simplest observable to calculate is the ratio of the number of jets with
a specified $E$ in the ensemble after quenching to that number in the
ensemble before quenching, a ratio that is the analogue in our model of $R_{\rm AA}^{\rm jet}$. 
In Fig.~\ref{fig:RAAplot} we show that if we choose $b=0.21$
we obtain a value for this ratio that is comparable to the measured
value of $R_{\rm AA}^{\rm jet}$~\cite{Raajet:HIN}. 
This value of $b$ is qualitatively consistent with what has been found in
other models.  For example, in the hybrid strong/weak coupling model
for jet  quenching, fitting the single model parameter
in that model to experimental measurements of $R_{\rm AA}^{\rm jet}$ yields the conclusion
that the thermalization length $x_{\rm therm}$ for jets is about 3-4 times
longer in QCD plasma with temperature $T$ than in strongly coupled ${\cal N}=4$ SYM
plasma with the same temperature~\cite{Casalderrey-Solana:2014bpa,Casalderrey-Solana:2015vaa}.

The blast wave temperature evolution is of course much simpler than
the true hydrodynamic evolution of the strongly-coupled plasma in
heavy ion collisions. In addition, the use of the metric (\ref{metric-1}) with
a $T$ that varies in space and time has
the downside that it neglects flow, viscosity, and gradients in the
plasma: (\ref{metric-1}) itself is the metric corresponding to a constant-temperature
plasma and when we insert a  $T$ that varies in space and time into it
what we obtain is not a solution to Einstein's equations for a plasma
whose temperature varies in space and time, meaning that it 
does not describe a solution to hydrodynamics: as noted, it neglects flow, viscosity and
gradients.
We have done brief 
and preliminary investigations where we have chosen more realistic
hydrodynamic backgrounds. 
First, we have tried taking the temperature evolution from the viscous hydrodynamic
simulation of a heavy ion collision in Ref.~\cite{Habich:2014jna} and used
the metric (\ref{metric-1})
for this $T$ that varies in space and time.
Second, we have tried taking both the temperature
and fluid velocity  from Ref.~\cite{Habich:2014jna} and implementing 
higher
order gradient corrections to the metric to include the effects of
flow and viscosity, as well as gradients in the fluid as in Ref.~\cite{Rajagopal:2015roa}.
From this brief study,
it appears that the alternative temperature profiles have only small
quantitative effects on the energy loss, while the presence of flow
and gradients may have somewhat larger effects but effects that are still only
quantitative, not qualitative.
Because of the computational complexity of such calculations, we
postpone the inclusion of a full hydrodynamic background including
fluid velocity, viscosity and gradients in the analysis of the full
ensemble of jets to future works.
In our limited study, including flow in the plasma profile appears
to decrease the energy loss. Improving upon our oversimplified 
treatment (aka neglect) of energy loss before hydrodynamization and
after hadronization would increase the energy loss; we leave these
investigations to future work also.

Before we turn to our results, we note that we shall be comparing our
results for how various observables are modified by passage through
the plasma to CMS measurements of jets in PbPb collisions from which
the effects of the jet energy resolution of the detector have not been
unfolded.  This means that, as described
in Refs.~\cite{Chatrchyan:2012nia,Chatrchyan:2013kwa},  
the appropriate baseline against which
to compare these measurements 
is not data from proton-proton collisions {\it per se}, because
the 
jet energy resolution of the CMS detector differs in PbPb and proton-proton collisions.
What we need to do, therefore, is to take as inputs to our calculations
distributions appropriate for unsmeared proton-proton collisions 
(we shall use \pythia simulations
thereof), calculate the modifications to observables after we run our
ensemble of jets through a droplet of plasma, and then before comparing
to measured data we must smear the jet energy in  
our initial (proton-proton)
ensemble and in our final ensemble after quenching, in both cases
using the Gaussian smearing functions for 0-10\% centrality PbPb collisions
provided by the CMS collaboration
in Ref.~\cite{Chatrchyan:2012gt}.  In Figs.~\ref{fig:rhoplot} and \ref{fig:RAAplot} and in the following Section, 
we can then compare to CMS measurements
in PbPb collisions that have not been unfolded, and to the pp baseline against which
CMS compares these measurements, namely simulated proton-proton jets smeared
to take into account the difference in the jet energy resolution of the CMS detector in PbPb and
proton-proton collisions.

\section{Results, Conclusions and Outlook}

\label{sec:results}

In previous Sections, we have described our construction of an
ensemble of null strings in the dual gravitational description of ${\cal N}=4$
SYM theory that we shall use as a model for an ensemble of jets.
Using null strings means that we are automatically describing flows
of energy whose opening angles do not change, in vacuum,
making them natural as models for high energy jets in QCD whose
jet mass and jet energy do not change, in vacuum.
Building an ensemble in such a way that it can serve as a model for
an ensemble of jets produced in proton-proton collisions 
required us to use key further inputs from several different directions.
First, we chose the probability distribution for the energy and opening
angle of the jets represented by the strings in our ensemble from
perturbative QCD calculations of these quantities in proton-proton collisions.
This alone is not enough, however, as we must specify the distribution of
energy density along our null strings.  We have done so in a way that incorporates a 
striking regularity of the dynamics of strings in AdS that we identified in Section 2.
We found that a large class of strings that are initially not null, and in particular that
have a vanishing initial opening angle, evolve to become null,
and as they do the energy density distributed along them takes on a particular scaling 
form, when scaled relative to the downward angle of the string endpoint, which is proportional
to the opening angle of the jet that the string models.
In Section 3 we have seen that if we choose the probability distribution for the opening
angles of the jets in our ensemble according to the results of a perturbative QCD
calculation, and then distribute the energy density along the string that
represents each jet according to the scaling form that we identified from our
analysis of string dynamics, we obtain an ensemble of jets with a mean jet shape
that is in excellent agreement with that measured in proton-proton collisions, see Fig.~\ref{fig:rhoplot},
upon fixing $a$, the first of two free parameters in our model.

One way of thinking about the way that we have chosen our ensemble of 
strings is that we have only included a subset of all possible null strings, the
subset whose energy is distributed along them such that they can be obtained
by starting with non-null strings whose initial opening angle vanishes initially
and letting these non-null strings evolve.  The string dynamics turns these initially
non-null strings into jet-like, null, strings with a particular form for their energy
density distribution.  Like any null string, these null strings can be created by
initializing them that way.  But, this subset of null strings can also be created by
starting with non-null initial conditions and letting the string evolve and nullify.
Although the property of the string dynamics that we are employing is 
striking, we have no first-principles argument for why we should choose this subset
of null strings.  Similarly, we do not know whether the only strings that can
reasonably be used as models for jets are strings that are null from the beginning.
We leave to future work constructing the bulk-to-boundary propagator for non-null strings
which nullify, computing the gauge theory stress-energy tensor at  early times, and investigating whether there are some initially non-null strings whose
opening angle is reasonably constant at all times, including before nullification.
With such a computation of the gauge theory stress-energy tensor 
at early times in hand, further investigations would also be possible, including trying to design initially non-null strings that have a reasonably constant opening angle at early times and that at the same
time have (fractal) substructure, as QCD jets do. 
For the present, we have an ensemble of strings that model an ensemble
of jets in vacuum with a mean jet shape as 
shown in Figure~\ref{fig:rhoplot}.

Next, we send this ensemble of strings through the simplified model for the expanding, cooling
droplet of plasma that we have described in Section 3.
There are many ways in which our treatment of the medium that our
ensemble of strings sees could be improved in future work.
As noted in Section 3, we have neglected any interactions between 
the strings and the medium before $\tau=1$~fm$/c$ and after the time
when the plasma cools below $T=175$~MeV.  Both these oversimplifications
can be revisited in future work.  Again as described in Section 3, we have used a blast wave model 
for the dynamics of the expansion and cooling of the droplet of plasma; this
can be revisited too.  At the same time in future when a full relativistic viscous hydrodynamic
treatment of the plasma is used instead of a
blast wave model, the AdS black hole metric should be augmented to include
the effects of shear viscosity and of fluid gradients.
For the present, we have the blast wave model for the expanding, cooling droplet
of plasma, as described in Section 3.  There, we have fixed the parameter $b$ (which is
the proportionality constant between the temperature of the QCD plasma
and the temperature of the ${\cal N}=4$ SYM plasma --- with its greater number of degrees of freedom ---
that we are using to model the QCD plasma).  The value of $b$ that we find corresponds to choosing the ${\cal N}=4$ SYM plasma to be between 3 and 4 times cooler than the QCD plasma that we are modeling, consistent with other estimates made in quite different ways~\cite{Casalderrey-Solana:2014bpa,Casalderrey-Solana:2015vaa}.

\begin{figure}[t]
\centering %
 \includegraphics[width=12cm]{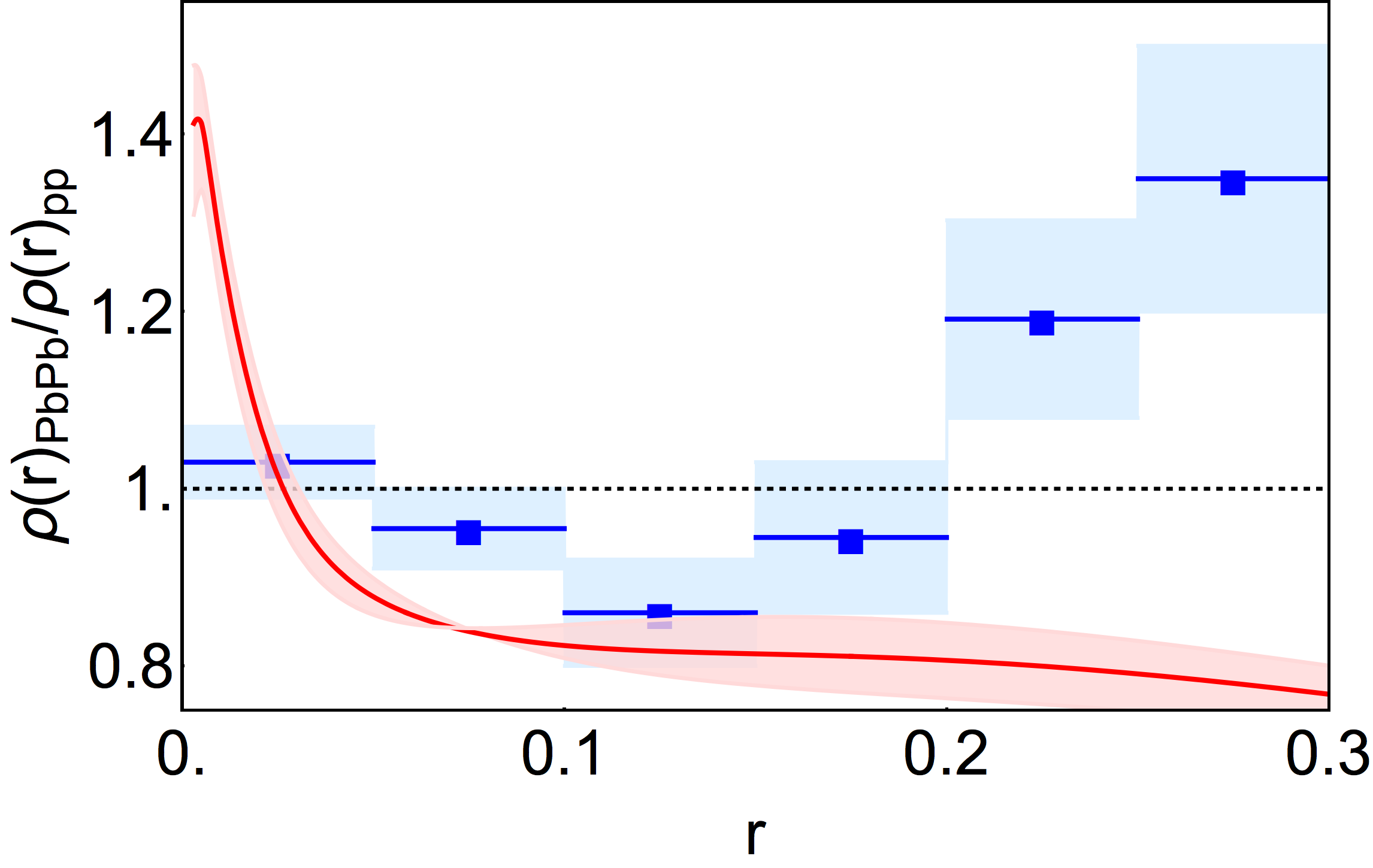} \caption{The 
 ratio of the mean shape of the jets in our ensemble after they have been quenched by
 their passage through the expanding cooling droplet of plasma, 
 as described in Section~\ref{sec:model},
 to their mean shape before quenching, shown in Fig.~\ref{fig:rhoplot}.
As in Figs.~\ref{fig:rhoplot} and \ref{fig:RAAplot}, the pink band shows
our results for $a=1.8-2.5$ with $a=2$ shown in dark red. 
The CMS measurement~\cite{Chatrchyan:2013kwa} 
of the modification of the mean jet shape, namely the ratio of its 
measured value for jets with energy above 100 GeV reconstructed with anti-$k_t$ $R=0.3$ in 
 0-10\% central PbPb collisions with $\sqrt{s_{NN}}$=2.76~TeV  to that in proton-proton collisions, 
as in Fig.~\ref{fig:rhoplot}, is shown as the blue symbols.
The discrepancy between our calculation and the experimental measurements 
at large angles $r$ relative to the center of the jet is
expected, since we do not include any analogue of the 
particles originating from the wake that the jet leaves behind in the plasma,
some of which must necessarily 
be reconstructed as a part of the jet in any experimental
analysis even after background subtraction~\cite{Casalderrey-Solana:2016jvj}.}
\label{fig:modrhoplot} %
\end{figure}


After sending our ensemble of jets through the droplet of expanding, cooling plasma,
we calculate the new mean jet shape in the ensemble, after quenching.
Note that, as in any experimental analysis, we impose 
a cut on the jet energy. In order to compare our results for the modification of the
mean jet shape to the experimental measurements in 
 Ref~\cite{Chatrchyan:2013kwa},
we impose the cut $p_T^{\rm jet}>100$~GeV, meaning
that any jet whose transverse momentum drops below 100 GeV 
upon propagation through the plasma is removed from the ensemble.
This means that even though every jet in the ensemble gets wider
as it propagates through the plasma, because jets that are initially wider
lose more energy and hence are more likely to drop below 100 GeV the mean
opening angle of the jets in the ensemble can decrease~\cite{Rajagopal:2016uip}.
Indeed, when we plot the ratio of the mean jet shape
of the jets in our ensemble after quenching to that in our ensemble before quenching
in Fig.~\ref{fig:modrhoplot} 
we see that quenching makes the mean jet shape get narrower in our calculation.
In Fig.~\ref{fig:modrhoplot} we compare the results of our calculation to the CMS data of Ref.~\cite{Chatrchyan:2013kwa}, finding qualitative agreement at
small $r$, namely close to the core of the jet, where we see that the ratio plotted in the figure
drops below one in our calculation and in data. 
This confirms that even though every jet in the ensemble gets wider as it propagates
through the plasma, the mean jet shape of the jets in the ensemble with
$p_T>100$~GeV gets narrower.
At larger $r$, our model does not include the soft particles
coming from the wake in the plasma --- which carry the momentum
lost by the jet and which therefore must
contribute to the reconstructed jet~\cite{Casalderrey-Solana:2016jvj}.
Including these effects  increases the number of soft
particles at all angles in the jet cone, which pushes the ratio plotted in Fig.~\ref{fig:modrhoplot} significantly upwards at 
larger $r$~\cite{Casalderrey-Solana:2016jvj}.


\begin{figure}[t]
\centering
 \includegraphics[width=12cm]{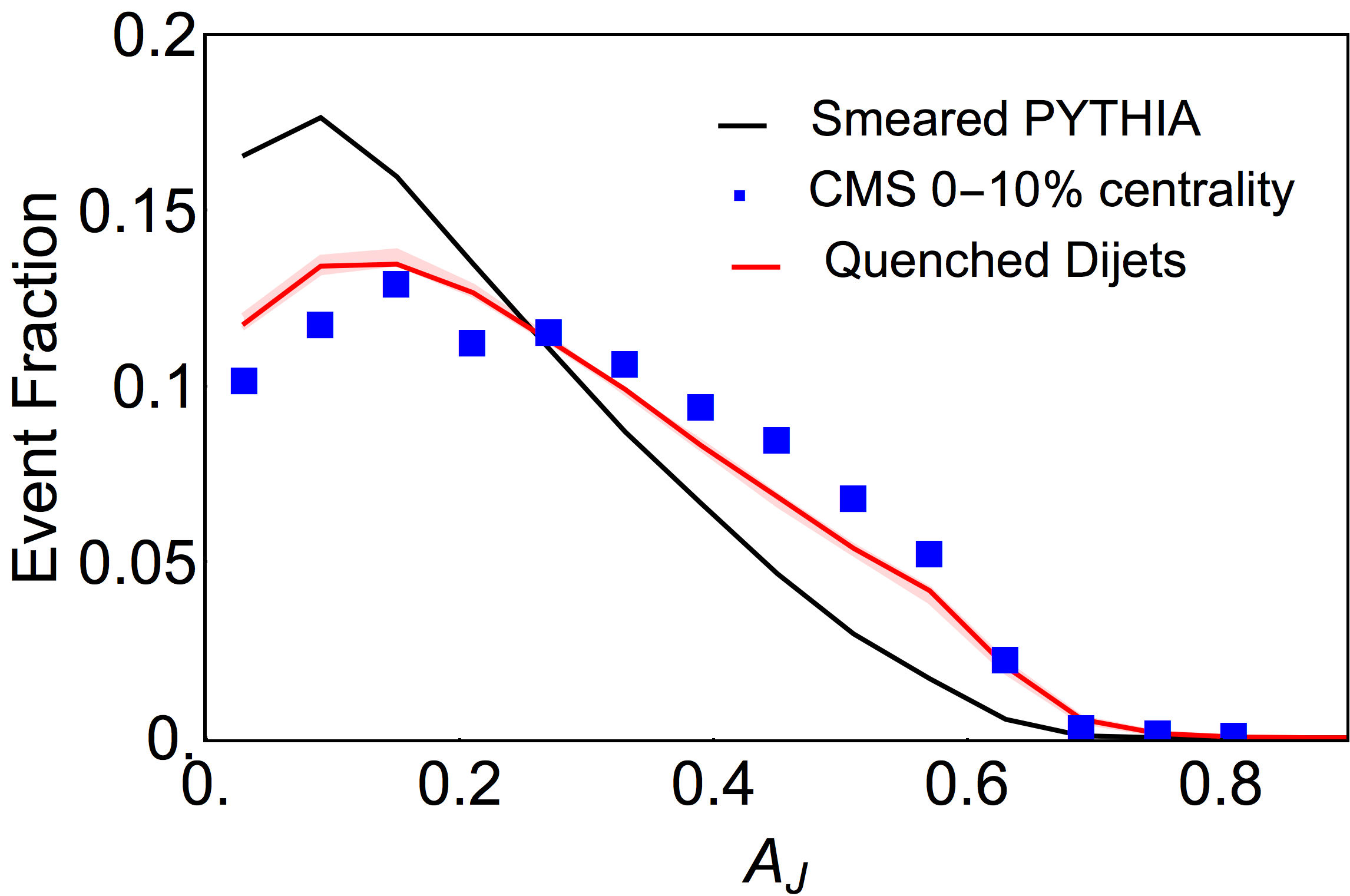} \caption{
 The dijet asymmetry distribution in our ensemble of holographic jets, before quenching (black curve)
 and after
 propagation through the strongly-coupled plasma of Section \ref{sec:model} (red curve).
The red curve is drawn for $a=2$, with the pink band indicating $a=1.8-2.5$.
The CMS measurement of the dijet asymmetry distribution in the
$\SIrange{0}{10}{\percent}$ most central Pb\textendash Pb collisions with $\sqrt{s_{NN}}=2.76$~TeV
from Ref.~\cite{Chatrchyan:2012nia} is shown in blue symbols.
In order to make a comparison to this data, as we described at the end of Section 3 we smeared both the unquenched input distribution that we obtained from \pythia and the distribution after quenching that is the output of 
our calculation, doing the smearing  as described in 
Refs.~\cite{Chatrchyan:2012nia,Chatrchyan:2013kwa},
}
\label{fig:dijetmodplot} %
\end{figure}

A by now classic signature of the modification of jets in heavy ion
collisions due to their passage through the strongly coupled plasma is 
a significant enhancement  in the dijet asymmetry.
In events in which at least two jets are reconstructed, the dijet asymmetry
is defined as $A_{J}\equiv (p_{T,1}-p_{T,2})/(p_{T,1}+p_{T,2})$,
where $p_{T,1}$ and $p_{T,2}$ are the transverse momenta of the
jets with the largest and second-to-largest transverse momenta.
The $A_J$ distribution is reasonably broad already in proton-proton
collisions, see the black curve in Fig.~\ref{fig:dijetmodplot}.
The two jets in a dijet need not be back-to-back and need not have
the same energy first of all because there may be three or more jets in
the event and second of all because of the interplay between the substructure
of jets and the algorithms via which jets are found and reconstructed.  In heavy ion collisions,
jet quenching introduces a significant new source
of dijet asymmetry, since one jet in the dijet will always lose more energy
as it traverses the plasma than the other.  A further broadening
of the $A_J$ distribution is thus a signature of jet quenching.

The $A_J$ distribution in the absence of quenching cannot be captured fully in our model, since we have no analogue of
jet finding or reconstruction and since we have no events with more
than two jets.
What we have done is to construct an ensemble of back-to-back dijets whose
$A_J$ distribution is as in proton-proton collisions, taking 
that input distribution  
from \pythia, and using a half of one of our strings to represent each
jet in a dijet pair. 
We have constructed an ensemble of roughly five
million dijet events with a distribution of asymmetries, in addition
to the distributions of opening angles, energies, starting positions,
and directions within the plasma as in the computation of the jet
shape modification.
The $A_J$ distribution from this ensemble, with the jet energies 
suitably smeared to take account of the jet energy resolution
of the CMS detector, as we described at the end of Section 3,
is
shown as the black curve in Fig.~\ref{fig:dijetmodplot}.
Note that although in the analysis reported in Fig.~\ref{fig:dijetmodplot}
we have only used back-to-back dijets, we have also constructed
a (smaller) ensemble in which 
we have not assumed that the jets are back-to-back, instead taking
 the distribution of angles
between the jets in dijet events from that measured in Ref. \cite{Chatrchyan:2012nia}.
This distribution is peaked, favoring jets which are back-to-back, and we have
checked that our results for the modification of the distribution of the dijet
asymmetry $A_J$ are not significantly different in this case than in our (larger)
ensemble that is the basis for Fig.~\ref{fig:dijetmodplot},
in which the dijets are always formed from a back-to-back pair.  What is important
is the choice of initial $A_J$  distribution.

Next, we send each dijet in the ensemble through its droplet of strongly coupled plasma.
Following Ref.~\cite{Chatrchyan:2012nia}, we then smear the jet energies
as described at the end of Section 3, select those events in which
$p_{T,1}>\SI{120}{\GeV/c}$
and $p_{T,2}>\SI{30}{\GeV/c}$ after quenching,
and compute the $A_J$ distribution for this ensemble of 
dijets that have been quenched
via their propagation through the plasma. 
Our results are shown in red in Figure~\ref{fig:dijetmodplot}, compared
with CMS data from the $\SIrange{0}{10}{\percent}$ most central PbPb collisions. 

We find qualitative agreement between the modification to the distribution of the
dijet asymmetry $A_J$ 
computed in this simple holographic model and heavy ion collision data from CMS. 
We anticipate that the largest systematic effect
not represented in the pink band in Figure~\ref{fig:dijetmodplot}
arises from the absence of three-jet events in our calculation, since
these are in fact the origin of much of the dijet asymmetry in proton-proton
collisions. We leave the construction of an ensemble of holographic three-jet events
to future work.

%
%

\section*{Acknowledgments}

We thank Francesca Bellini, Jorge Casalderrey-Solana, Paul Chesler, David Chinellato, 
Yang-Ting Chien,
Andrej Ficnar, Alexander Kalweit, Yen-Jie Lee, Simone Marzani, Chris McGinn, Guilherme Milhano, Daniel Pablos and
Jesse Thaler for useful discussions. We are especially grateful to
Simone Marzani for providing the formulas of Ref.~\cite{Larkoski:2014wba}.
KR acknowledges the hospitality of the CERN Theory Group.
AS is partially supported through the LANL/LDRD Program.
This work is supported by the U.S. Department of Energy under grant
Contract Number DE-SC0011090.

\bibliographystyle{bibstyle}


\bibliography{references}

\end{document}